\documentclass[12pt,preprint]{aastex}

\shorttitle{V471~Tau}
\shortauthors{Kami\'nski et al.}

\begin{document}

\title{MOST photometry and DDO spectroscopy of the \\ 
eclipsing (white dwarf + red dwarf) binary  V471~Tau\footnote{Based 
on data from the MOST satellite, a Canadian Space Agency mission 
jointly operated by Dynacon Inc., the University of Toronto Institute for 
Aerospace Studies and the University of British Columbia, with the 
assistance of the University of Vienna, and on data obtained at the 
David Dunlap Observatory, University of Toronto.}}

\author{Krzysztof Z. Kami\'nski}
\affil{Astronomical Observatory, Adam Mickiewicz University \\
ul. S{\l}oneczna 36, 60-286 Pozna\'n, Poland}
\email{chrisk@amu.edu.pl}

\author{Slavek M. Ruci\'nski}
\affil{David Dunlap Observatory, University of Toronto \\
P.O.~Box 360, Richmond Hill, Ontario, L4C~4Y6, Canada}
\email{rucinski@astro.utoronto.ca}

\author{Jaymie M. Matthews, Rainer Kuschnig, Jason F. Rowe}
\affil{Department of Physics \& Astronomy, University of British Columbia \\
6224 Agricultural Road, Vancouver, B.C., V6T~1Z1, Canada}
\email{(matthews,kuschnig,rowe)@astro.ubc.ca} 

\author{David B. Guenther}
\affil{Institute for Computational Astrophysics, 
Department of Astronomy and Physics, \\ Saint Marys University, 
Halifax, N.S., B3H~3C3, Canada}
\email{guenther@crux.stmarys.ca}

\author{Anthony F. J. Moffat}
\affil{D\'{e}partment de Physique, Universit\'{e} de Montr\'{e}al \\ C.P.6128, 
Succursale: Centre-Ville, Montr\'{e}al \\ QC, H3C~3J7, 
Observatoire du Mont M\'egantic, Canada}
\email{moffat@astro.umontreal.ca}

\author{Dimitar Sasselov}
\affil{Harvard-Smithsonian Center for Astrophysics \\ 60 Garden Street, 
Cambridge, MA 02138}
\email{sasselov@cfa.harvard.edu}

\author{Gordon A. H. Walker}
\affil{1234 Hewlett Place, Victoria, BC V8S 4P7, Canada}
\affil{Department of Physics \& Astronomy, University of British Columbia \\
6224 Agricultural Road, Vancouver, B.C., V6T~1Z1, Canada}
\email{gordonwa@uvic.ca}

\and

\author{Werner W. Weiss}
\affil{Institut f\"{u}r Astronomie, Universit\"{a}t Wien\\
T\"{u}rkenschanzstrasse 17, A-1180 Wien, Austria}
\email{weiss@astro.univie.ac.at}

\begin{abstract}
The Hyades K2V+WD system 471~Tau is a prototype post-common 
envelope system and a likely cataclysmic binary progenitor. 
We present 10 days of nearly continuous optical photometry by 
the MOST (Microvariability \& Oscillations of STars) 
satellite and partly simultaneous optical spectroscopy 
from DDO (David Dunlap Observatory) of the binary.
The photometric data indicate that the spot coverage of the 
K dwarf component was less than observed in the past, 
suggesting that we monitored the star close to a 
minimum in its activity cycle.  Despite the low spot activity, 
we still detected seven flare-like events whose 
estimated energies are among the highest ever observed 
in V471~Tau and whose times of occurrence do 
not correlate with the binary orbital phase.  
A detailed $O-C$ analysis of the times of eclipse over the last 
$\sim35$~years reveals timing variations which could 
be explained in several ways, including perturbations by
an as-yet-undetected third body in the system or by
a small orbital eccentricity inducing slow apsidal motion.
The DDO spectra result in improved determinations 
of the K dwarf projected rotation velocity, $V_K \sin i = 92$ 
km~s$^{-1}$, and the orbital amplitude, $K_K=150.5$ km~s$^{-1}$.  
The spectra also allow us to measure changes in 
$H\alpha$ emission strength and radial velocity (RV) 
variations. We measure a larger $H\alpha$ velocity amplitude
than found previously suggesting that
the source of the emission in V471~Tau 
was less concentrated around the sub-white-dwarf 
point on the K star than had been observed in previous studies.
\end{abstract}


\section{INTRODUCTION}
\label{intro}

V471~Tau is a close eclipsing binary star ($V \sim 9$) 
consisting of a hot white dwarf and a red dwarf with  
an orbital period of 0.521 d. It is a member of the Hyades 
cluster \citep{werner97} and -- very likely -- a 
cataclysmic binary progenitor \citep{still03}. V471~Tau 
has been the subject of numerous investigations over 
the past 35 years; cf.\ the main contributions by 
\citet{skillman88}, \citet{clemens92}, \citet{obrien01}, 
\citet{ibanoglu05} and \citet{hussain06}.  

The V471~Tau system may be the prototype of a post-common 
envelope binary with a white dwarf and a main sequence star.  
The mass and radius of both components can be measured
with high accuracy, while the K dwarf which is spun up
to high rotation rates by tidal forces may be an analogue 
for rapidly rotating pre-ZAMS stars like AB~Dor. 
Simultaneous precise time-resolved photometry and 
spectroscopy which cover phases of eclipse in the 
V471~Tau system can sample the spot coverage of the 
K dwarf.  Eclipse timing can measure apsidal 
motion in the binary and test whether the system is 
actually a triple one with a third undetected component.
We therefore organized a coordinated campaign of spacebased 
photometry from the MOST satellite and 
groundbased spectroscopy from DDO.

We present new MOST and DDO observations of V471~Tau
in Section~\ref{observations}. The MOST light curve and its
changes are discussed in Section~\ref{interpret-lc}
while Section~\ref{interpret-sp} gives a description
of the spectroscopic observations. Section~\ref{summary}
summarizes the combined results.

\section{OBSERVATIONS OF V471 TAU}
\label{observations}

\subsection{MOST photometry}
\label{photometry}

The MOST (Microvariability \& Oscillations of STars) 
space mission \citep{walker2003, matthews2004} was 
designed to perform high-precision optical photometry 
of bright stars with long time coverage and high 
duty cycle. MOST is equipped with a 15-cm telescope and a 
custom broadband filter (spectral transmission 
peak $\sim 5500$~\AA\ and FWHM $\sim 3000$~\AA). 
The polar Sun-synchronous orbit of the satellite 
allows it to monitor stars in the anti-solar direction 
for up to 60 days without interruption.

MOST observed V471~Tau for 10.0 days during 4 -- 14 December 2005 
(in Terrestial Time Julian Days: 
2,453,708.5117 -- 2,453,718.5122, see below in
Section~\ref{timing}), covering just over 
19 orbital periods of the binary system.  
The photometry was obtained in MOST's Direct Imaging mode 
\citep{rowe06}, with a slightly defocused stellar 
image sampled in a CCD sub-raster. The exposure time was 6.52 s, 
sampled at 10-s intervals. Two reference stars in the same field 
(GSC~01252-00692, $V = 8.9$ and GSC~01252-00046, $V = 9.8$)
were observed simultaneously in the same way
to calibrate instrumental or satellite orbital artifacts.

The MOST instrument focal plane can be illuminated by 
scattered Earth light whose level is modulated by 
the MOST orbital period of $P_M \simeq 101$ min.  
The amplitude and complexity of the stray light 
background variations depend on the season of observing, 
the location of the target star relative to the 
bright limb of the Earth and the orientation (roll) of the 
spacecraft.  In the case of the V471~Tau photometry, 
the periodic fluctuations in background translated into 
photometric uncertainties in the stellar signal ranging 
from point-to-point scatter with $\sigma \simeq 0.003$ 
(about 3 mmag) at stray light minimum to a point-to-point
scatter of about $\sigma \geq 0.1$ at stray light maximum.

The dark and flat field corrections were performed
by monitoring individual pixel responses during
test exposures on fields empty of stars bright enough
to rise above the background.  Photometry was extracted
from the stellar images using a Moffat-profile
point spread function model \citep{moffat69}.
The correlation in the raw photometry
between the instrumental magnitude light
curve and the estimated sky background was removed
as described in \citet{rowe06}. About 29\%
of the total number of data points were rejected because of
pixel saturation during phases of the highest stray light
in the MOST orbit and high cosmic ray fluxes
when MOST passed through the South Atlantic Anomaly,
as indicated by the orbital model of the local magnetic
field strength. Additionally, about 6\% of data
points were rejected because of the relative uncertainty
exceeding $\sigma=0.015$ of the mean light level.

The reduction and selection procedure left 56,383 measurements
containing gaps of variable
length spaced roughly by the MOST orbital period,
resulting in a net  duty cycle of about 65\%.
(We later conducted a period search
after an even stricter selection of the data, 
with a duty cycle of 59\%, as described in
Section~\ref{periodograms}.)  The time sampling and duty 
cycle provide excellent coverage in binary
orbital phase and during the eclipses of V471~Tau.
Note that the orbital period of the binary of close
to 1/2 day always created a phase-coverage problem for
ground based observations; the MOST data are entirely free of
this limitation.
The MOST photometry data (raw, and the reduced light curve 
used for analysis in this paper) are available 
in the MOST Public Data Archive on the Science page of the 
MOST web site: www.astro.ubc.ca/MOST.


\subsection{V471~Tau light curve}
\label{lc}

\begin{figure}    
\epsscale{1.0}
\plotone{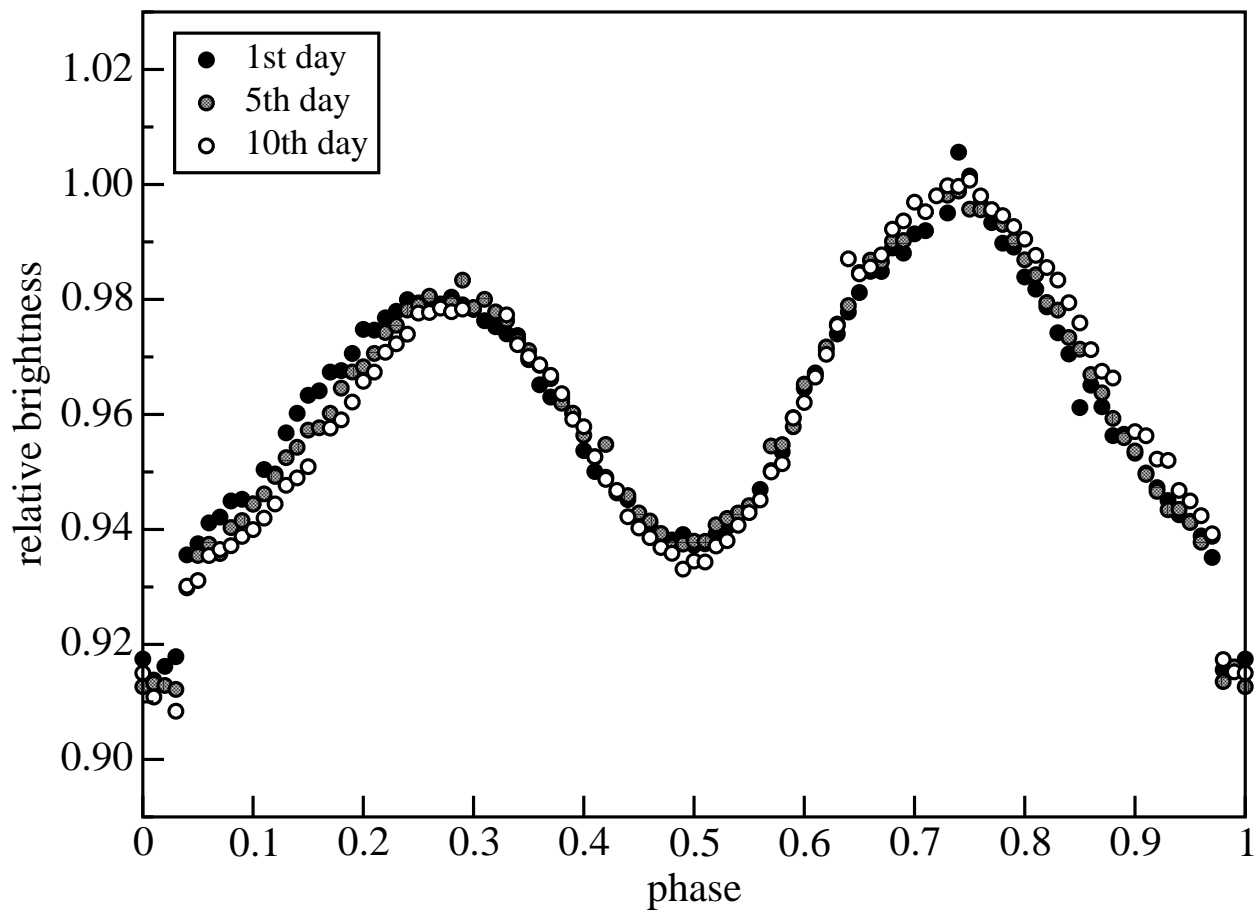}
\caption{The mean V471~Tau light curve,
averaged in phase with 0.01 phase bins,
for 3 selected days of the MOST observations
at the beginning, middle and end of the run.
A lack of obvious changes in the light curve during our
observations beyond the global shifting at phases $\sim\ 0.75\ -\ 1.25$,
can be interpreted as a relatively low activity
in the spot re-arrangement. \label{fig1} 
}
\end{figure}

The 19 orbital cycles of the binary monitored by MOST allowed 
us to investigate changes in the light
curve from cycle to cycle, which is normally interpreted as 
migration and evolution of spots on the 
magnetically active K dwarf component \citep{ibanoglu78}.  
The MOST data were divided into 1-day long subsets 
and those subsets were phased with the known orbital period 
of V471~Tau.  Three of these subsets, from the beginning, middle 
and end of the 10-day run, are overplotted in Figure~\ref{fig1}. 
A subtle trend is visible in that the rising portion of the 
light curve (in the phase interval 0.05 -- 0.25) moves 
systematically later in phase with time, by a total of about 
0.04 over 10 days. There is some evidence of 
this shift during the falling portion of the curve in the 
phase interval 0.75 -- 0.95, but it is less pronounced. 
No phase shift is seen in the phase range 0.3 -- 0.7, 
within the photometric scatter.

\placefigure{fig1}

The changes seen in the MOST photometry resemble the 
``photometric wave migration'' first reported by 
\citet{ibanoglu78} and discussed below in 
Section~\ref{spot}. The average shift of the wave of 
$\sim 0.002$~phase/day indicates that it would 
take $500 \pm 250$~days for the wave to make a full 
revolution ($P_{migr}$). This is somewhat longer than the 
wave migration period found in previous studies
(from $\sim 180$ d by \citet{ibanoglu89} to $372$ d 
by \citet{skillman88}), although our estimate of the 
drift rate, based on only 19 orbital cycles, is necessarily crude.

\begin{figure}    
\epsscale{1.0}
\plotone{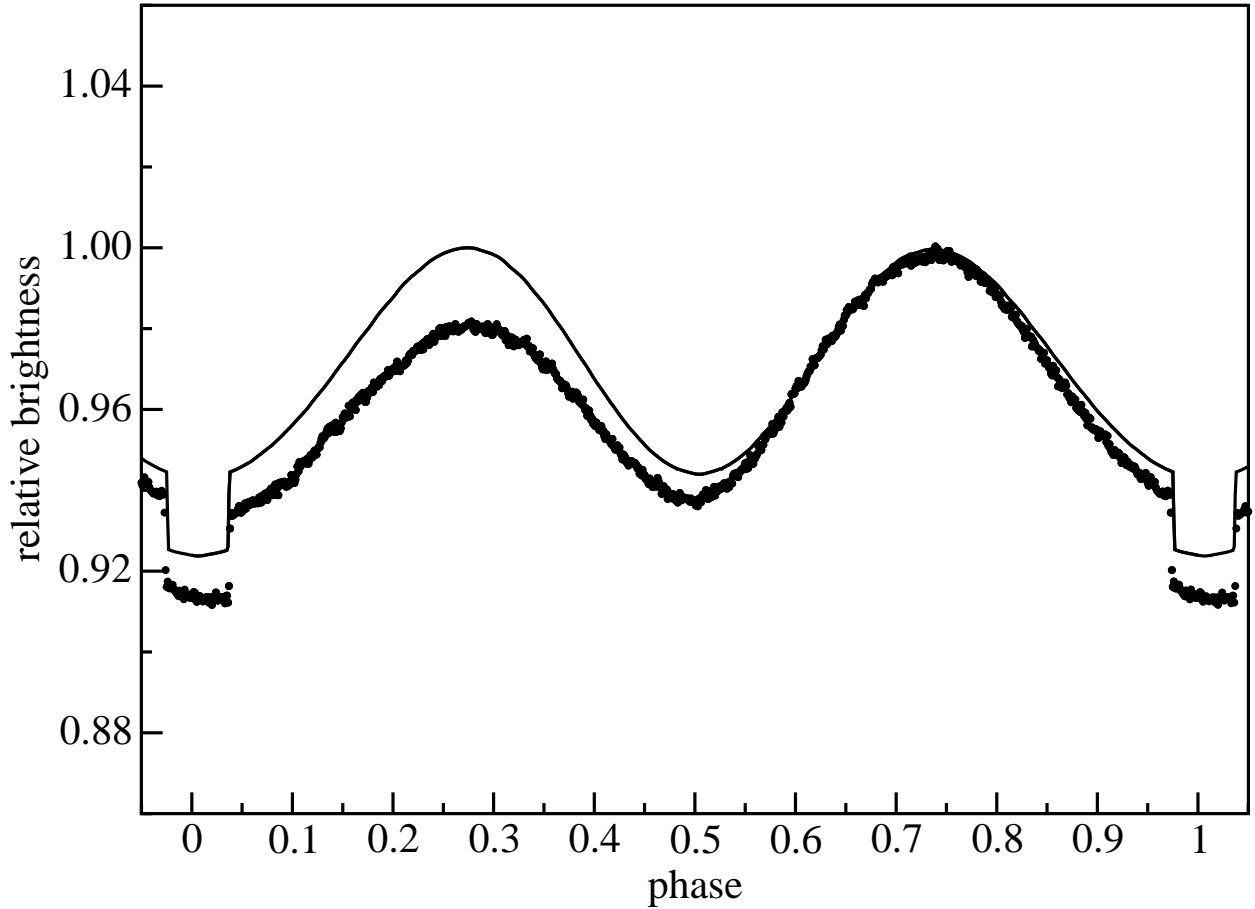}
\caption{The mean V471~Tau light curve averaged in phase with
0.001 phase bins (dots).
The line shows the light curve calculated
by the PHOEBE package \citep{prsa05},
based on the published V471~Tau parameters.
Deviations caused by the spots on the K~dwarf are not
included in the model. \label{fig2}
}
\end{figure}

Seeing that the systematic changes in the light 
curve during the 10-day span of our observations were 
relatively small, with apparent shifts less than $0.01$ 
mag at a given orbital phase, we calculated a mean
light curve from the entire time series.  
This is presented in Figure~\ref{fig2}.

\placefigure{fig2}

\subsection{DDO spectroscopy}
\label{spectroscopy}

\begin{figure}    
\epsscale{0.85}
\plotone{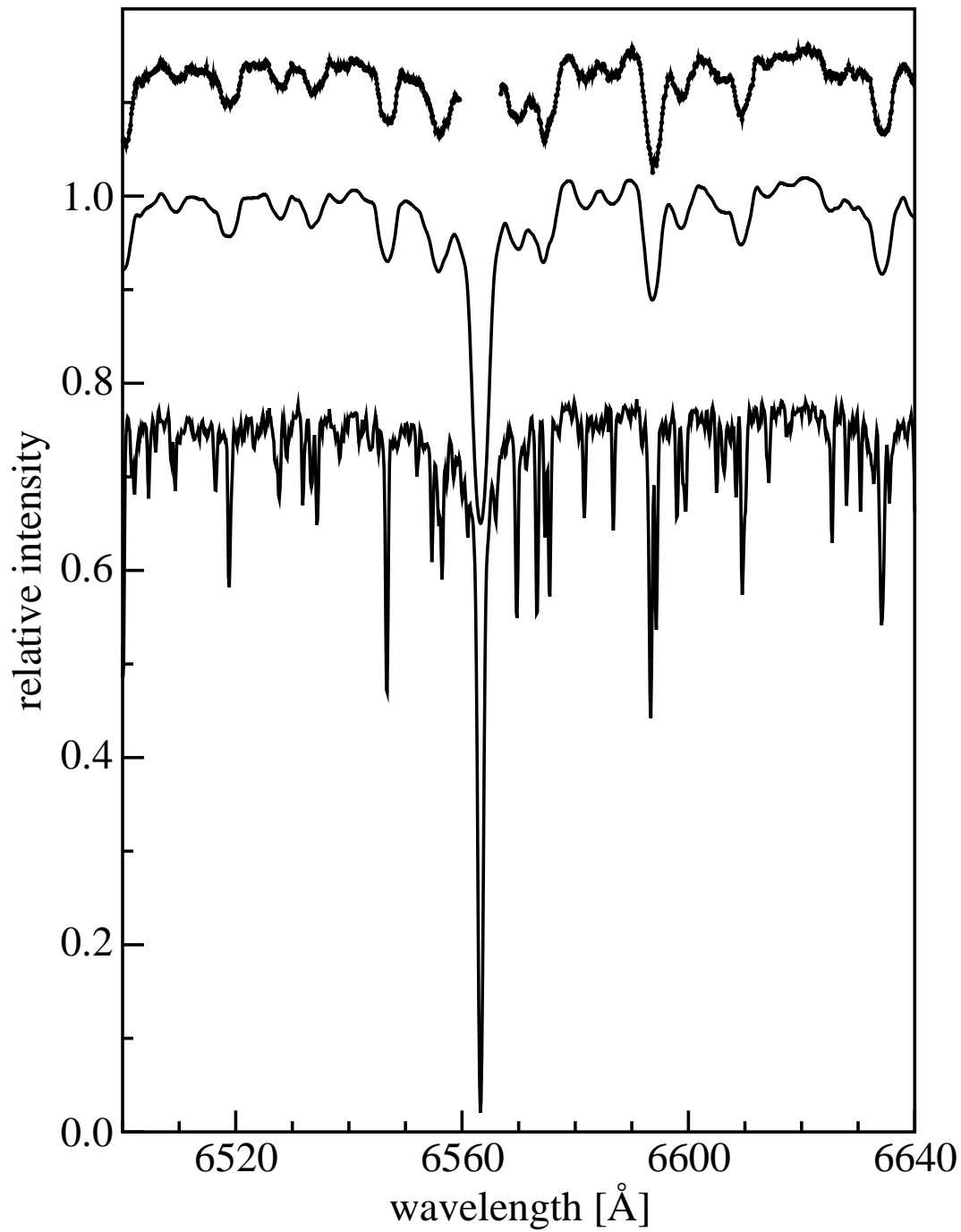}
\caption{Comparison of the DDO spectra of V471~Tau.
Top: The average of the 27 best quality spectra after 
correction shifts for the orbital motion. The $H\alpha$ line was omitted
due to its variability (see Subsection~\ref{halpha}) Middle: 
The standard star HD~3765 spectrum after convolution 
with the broadening profile. Bottom: The HD~3765 spectrum as
observed. \label{fig3}
}
\end{figure}

We obtained ground based spectroscopy of V471~Tau which 
partially overlapped with the MOST photometric
run during 7 -- 19 December 2005 (see Table \ref{tab1}).  
A total of 37 spectra at a dispersion of 
0.14~\AA/pixel were collected using the Cassegrain 
spectrograph of the 1.88-m telescope at the David 
Dunlap Observatory. Since we expected the K-type 
dwarf in the system to dominate the flux at 
optical wavelengths, the wavelength range of the spectra 
was centered at $H\alpha$ line, covering a red spectral window between
6425 and 6705~\AA,\ (Figure~\ref{fig3}).  
This region contains a multitude of telluric lines which
were removed during standard reductions performed 
using IRAF\footnote{IRAF is distributed by the National 
Optical Astronomy Observatories, which are operated by 
the Association of Universities for Research in 
Astronomy, Inc., under cooperative agreement with the 
National Science Foundation.} routines. The spectra 
were taken with the integration times and at
intervals of about 30 minutes and could 
not cover all orbital phases of the binary because of the
night/day breaks, commensurability of the binary period
with one day and interruptions due to weather. The long
integration times preclude any use of the spectroscopic
data for improvement of the eclipse timing described in
Section~\ref{timing}. 

\placefigure{fig3}

\placetable{tab1}

\section{INTERPRETING THE LIGHT CURVE}
\label{interpret-lc}

The phase diagram of the mean light curve presented in 
Figure~\ref{fig2} was modeled using the 
PHOEBE software package \citep{prsa05}, based on the 
Wilson-Devinney model. The orbital and physical 
parameters of both stars in the system were adopted 
from \citet{obrien01}: $R_K=0.96 \,R_{\sun}$, 
$T_K=5,040 \, K$, $R_{WD}=0.0107 \, R_{\sun}$, 
$T_{WD}=34,500 \, K$, $a=3.30 \, R_{\sun}$, 
$i=77.4^\circ$; the subscripts $K$ and $WD$ 
signify the K and white dwarf components, respectively.
The atmospheric parameters for the red
dwarf component were set to typical values for a 
K dwarf; limb darkening = 0.52, 
gravity darkening = 0.32 and albedo = 0.5.

The resulting model reproduces the general nature 
and amplitudes of the double-wave variability, and 
the depth of the eclipse, seen in the MOST light curve, 
as shown in Figure~\ref{fig2}. 
It consists of the dominant smooth, wave-like 
variability and a relatively shallow (0.022-mag 
deep) total eclipse lasting 46.9 minutes, with steep shoulders 
each approximately 50 sec long. The photometric 
double wave is caused by ellipsoidal distortion of the 
K dwarf, with a minor modification due to the 
reflection effect. The asymmetry in the ellipsoidal distortion 
variability is believed to be due to spots on the K dwarf. 

\subsection{Spot coverage}
\label{spot}

\begin{figure}    
\epsscale{1.0}
\plotone{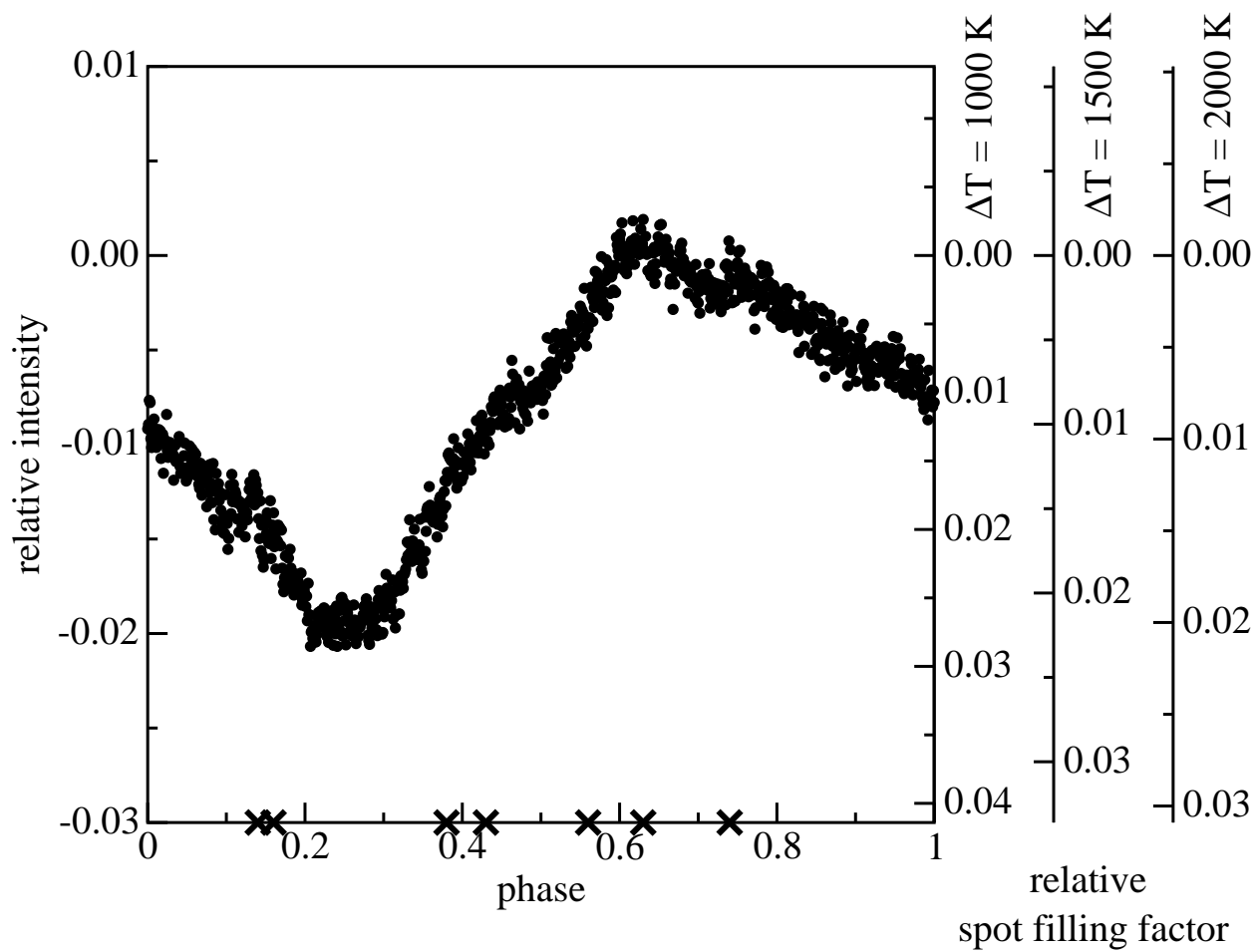}
\caption{Relative changes of the spot-filling factor versus the
orbital phase for different
spot temperatures, as indicated on the right vertical axis. 
The curve was obtained by comparing
observed light curve (averaged in phase with 0.001 phase bins)
with the theoretical one calculated with the PHOEBE package,
as described in the text.
The crosses on the bottom axis mark the phases of seven detected
flare-like events (see Subsection~\ref{flares}). \label{fig4}
}
\end{figure}

In order to study the locations and extent of spots 
on the surface of the K dwarf, we used the residuals 
between the observed light curve and the modeled 
light curve (Figure~\ref{fig4}) to estimate the required 
changes of the spot filling factor with orbital phase. 
Because of the orbital inclination of $77\degr$, it is only 
possible to estimate changes in the mean spot coverage 
on the K dwarf disk within the latitude range of
$-77\degr$ to $+77\degr$.  Recent Doppler imaging observations
of \citet{hussain06} revealed that the K dwarf is rotating 
rigidly; this is confirmed by our determination of
$V_{K} \sin i$ (Section~\ref{rotation}). 
As our run duration was only 2.5 times longer than the
time span of the observation used by \citet{hussain06},
we expect any changes of filling factor at a given phase to reflect
spot rearrangement caused by the star activity rather than
the star differential rotation.  Also, any spot interpretation
can address only the part of the spot coverage
which is longitudinally asymmetric.

\placefigure{fig4}

Our results indicate that the smallest spot coverage 
occurred during the orbital phases $0.6 - 0.7$,
while the largest occurred during $0.2 - 0.3$. We seem 
to have observed a totally different level of activity 
in the K dwarf than seen during the Doppler 
imaging observations by \citet{ramseyer95} and 
\citet{hussain06}. Our estimate of the peak-to-peak amplitude 
of the spot filling factor, $0.02 - 0.03$ 
(depending on the assumed spot temperature differential 
values of $\Delta T = 2,000 - 1,000 \,K$, as shown 
in Figure~\ref{fig4}), is many times smaller than the 
changes of $\sim 0.15$ observed by \citet{hussain06}
in November 2002. Also, in our data, the maximum spot 
coverage is inferred close to orbital phase 0.25,
while \citet{hussain06} found the maximum around phase 0.07. 
The {\it evolution\/} of the spot coverage during
the 10-day MOST observing run was still smaller, 
typically at a level of $\leq 0.01$, depending on the phase.

\subsection{Eclipse timing}
\label{timing}

The relatively slow 10-sec photometric sampling rate 
(compared with the eclipse branch duration of 50 sec) and 
the temporal gaps left after the data selection made it 
impossible to measure times of individual eclipses 
accurate to a few seconds.  Instead, we calculated
the average eclipse time on the basis of a phased light 
curve of the entire time series to compare with earlier 
eclipse times in the literature. The phases were computed 
with the linear ephemeris given by \citet{guinan01}. 
Because previous eclipses have been observed over a long 
time span ($\sim 35$~yr) and the orbital period of the 
binary is short, we adopted a uniform time system of 
Heliocentric Julian Date based on the Terrestrial Time (HJED), 
as advocated by \citet{bastian2000}.

\begin{figure}    
\epsscale{1.0}
\plotone{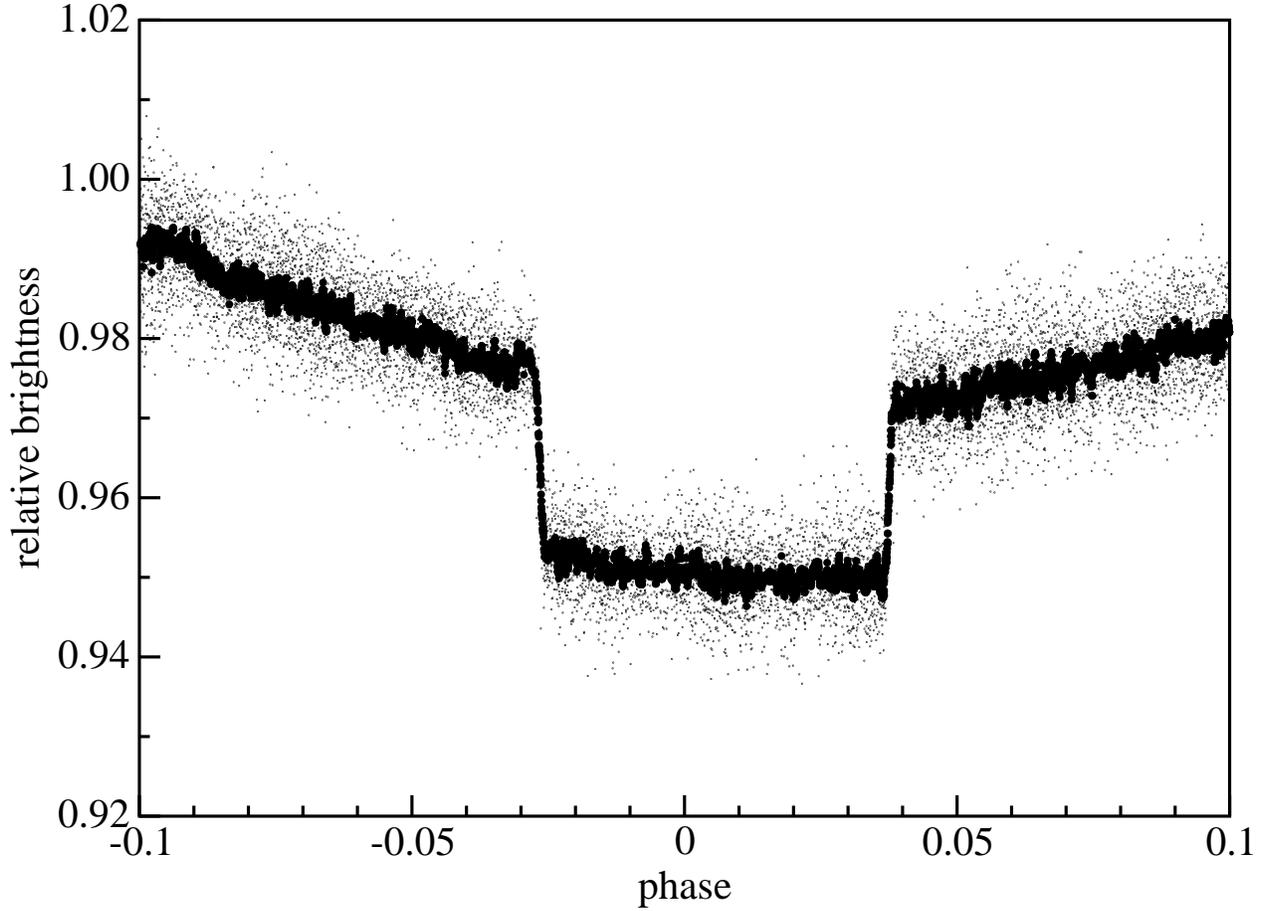}
\caption{All observations used for the V471 Tau eclipse 
timing (small dots) are shown in the phase diagram together 
with the running average data (large dots).
The phase shift of the mid-eclipse time relative
to the \citet{guinan01} linear ephemeris is clearly visible.
\label{fig5}
}
\end{figure}

\begin{figure}    
\epsscale{1.0}
\plotone{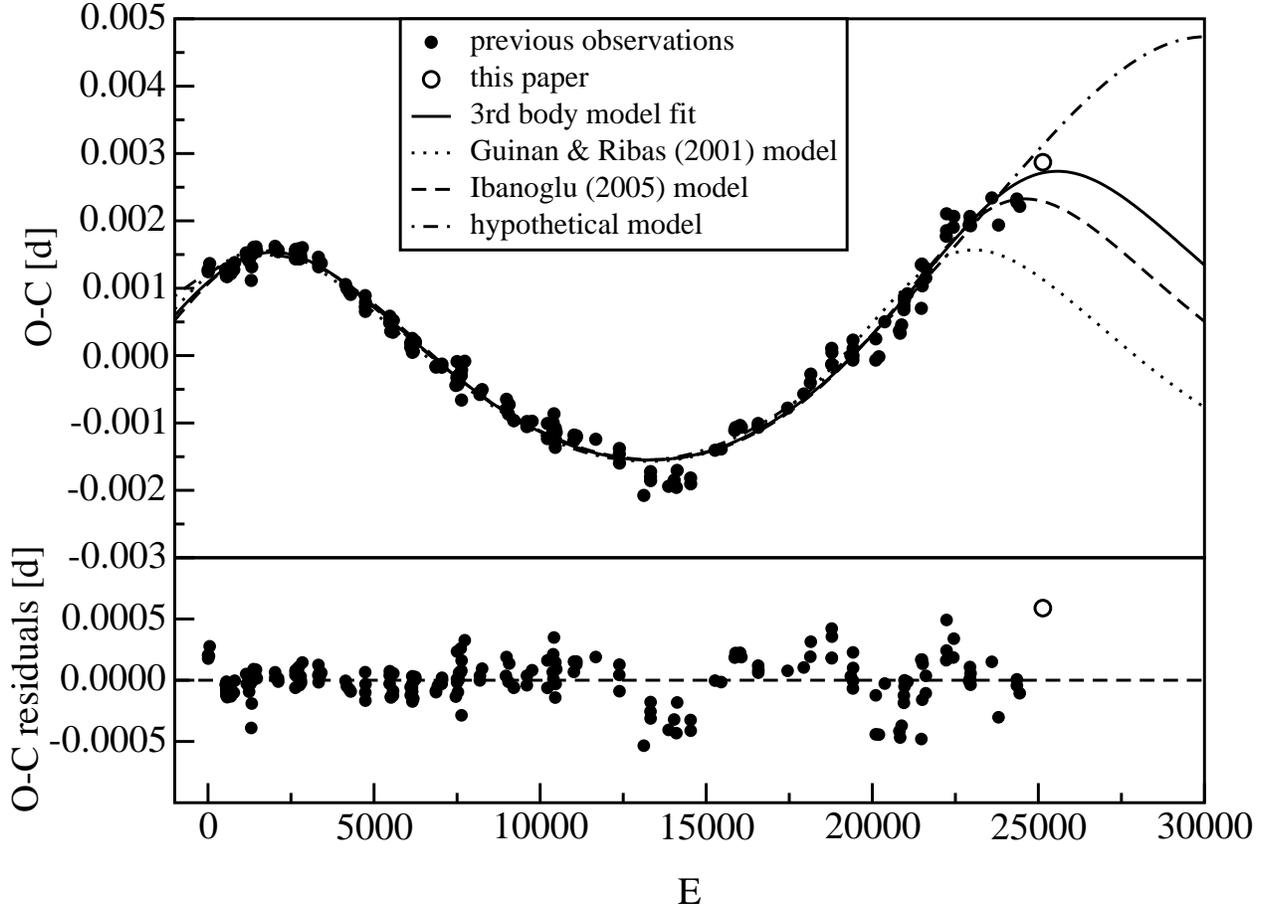}
\caption{The third-body model fits to eclipse timing observations 
of V471 Tau from the literature (filled circles).
The open circle is the new timing from the MOST observations.
A hypothetical, illustrative trend was created by adding a point of
$O-C=0.004480$ at $E=28,000$, that would follow the curve growing trend.
The bottom plot shows the residuals
of all available data with respect to the model
predictions of \citet{ibanoglu05}. \label{fig6}
}
\end{figure}

The eclipse time was determined after correction 
of the light curve for the local slope created by the 
photometric wave. Since all contacts of the eclipse
are not well defined (see Figure~\ref{fig5})
we determined the intersections of the averaged eclipse
branches with horizontal line at the mid-depth level.
The mid-point of both intersections corresponds
to the mid-point of the eclipse.
With the ephemeris of \citet{guinan01}, 
our mean epoch corresponds to $E=25,135$. The shift in 
the observed time of the mid-point of eclipse is large 
compared to the predicted zero phase by \citet{guinan01}: 
$O-C = +248 \pm 7$~seconds, or over 4~minutes 
(see Figure~\ref{fig5}). 
The MOST eclipse time determination is shown compared to 
all available published data (as discussed by 
\citet{ibanoglu05}) in Figure~\ref{fig6}. 
The $O-C$ curve continues an upward trend seen for about
the last 10,000 orbital cycles.  The implications of 
the MOST timing point are explored below.

\placefigure{fig5}

\placefigure{fig6}

\subsubsection{Third body}
\label{third}

The V471 Tau period changes visible in the eclipse $O-C$ 
diagram have been interpreted previously by several 
others as a light-travel-time effect caused by a 
perturbing third star in a long-period orbit in the system 
\citep{ibanoglu94,guinan01,ibanoglu05}. This explanation is 
attractive because it could be reconciled with the 
main features of the $O-C$ diagram.  It is also exciting 
because the mass of the hypothetical third body would  
be sub-stellar for a large range of possible inclination angles. 

Our new eclipse timing measurement shows that the 
long-anticipated downward bend in the $O-C$ diagram has 
not yet happened. Moreover, it deviates substantially from 
the most recent third-body model proposed  
\citep{ibanoglu05} by 52 sec, which is 3.6 times larger 
than $\sigma$ of the residuals for this model, as shown in 
the lower panel of Figure~\ref{fig6}. Indeed, the MOST 
point is the largest deviation from this model so far 
observed.

Therefore, we decided to recalculate the third-body model 
utilizing the same formalism as in \citet{ibanoglu05}.
With the new data augmented by the MOST result, the 
semi-amplitude of the $O-C$ variations, the third-body 
orbital period and its mass function are all slightly larger 
and longer than those given by \citet{ibanoglu05}; see 
Table~\ref{tab2} for the full set of fitted parameters. 
The third-body orbital fit, although formally appropriate, 
remains uncertain because we still do not see the bend 
in the $O-C$ curve. In fact, as is shown in Subsection~\ref{sudden} 
below, it is reasonable to assume that the 
period has been constant since $E \approx 15,000$,
i.e., over the last $\sim14$ years. However, if we continue 
to see a straight line in future extensions of the 
$O-C$ diagram, this will not necessarily exclude the 
third-body hypothesis. Figure~\ref{fig6} includes a 
fit to a third-body model whose orbit has an even longer 
period which can still match the observations.  Note 
that the orbital inclination range necessary to preserve 
the sub-stellar mass of the third body will decrease to 
a very small range of angles if the current linear trend 
in the $O-C$ deviations continues.

The suggested parameters of the hypothetical third body
in the V471~Tau system indicate that this object may
be detectable with modern infrared telescopes
or interferometers. With a larger mass function and 
a longer orbital period than in \citet{guinan01}, 
the separation and brightness
of the third body can be as large as 0.9 arc second and
K $\sim$ 13.3 mag; see Table~\ref{tab6} for predictions 
of the third body parameters for its 
different orbital inclinations.

\placetable{tab2}

\placetable{tab6}

\subsubsection{Apsidal motion}
\label{apsidal}

If the binary orbit is even slightly ellipsoidal, it 
may show a motion of the line of apses. This explanation was
mentioned by \citet{herczeg75} and \citet{skillman88}, but 
then dismissed as an unlikely cause for the changes
in eclipse times. We performed a least-squares fit of the $O-C$ 
curve with the first-order formula given by 
\citet{todoran72} and found that a very narrow range 
of eccentricity, $e = 0.0121 \pm 0.0006$ (with 98\% 
confidence level), is required to explain the latest 
$O-C$ results we have presented.  See Table~\ref{tab3} 
and Figure~\ref{fig7}.  Although the orbit is expected 
to circularize in a very close binary system like 
V471~Tau, our fit to a slightly non-zero eccentricity is 
surprisingly close to the one we find from our radial velocity 
orbital measurements (see Section~\ref{rv} below).

\placetable{tab3}

\subsubsection{Sudden period changes}
\label{sudden}

Without assuming anything about the actual nature of 
the $O-C$ changes, it may be argued that the curve is 
composed of a few straight-line segments, each corresponding 
to a constant period, and of relatively short 
intervals where abrupt period changes take place. The 
portions of the $O-C$ diagram from epochs $E \approx 
2,500$ to $10,500$ and from $E \approx 15,000$ onwards appear to 
be consistent with two different constant 
periods. Least-squares linear fits to both segments of 
the $O-C$ diagram yield periods of 
$0.52118305(4)$ and $0.52118384(4)$ days, 
respectively (the errors in parentheses are given in units
of the last decimal place), corresponding to a relative 
period change of $\Delta P/P \simeq 1.5 \times 10^{-6}$.

A sudden period change may be explained as a result
of mass transfer or mass loss in a binary.
For V471~Tau, we do not know if the possible donor,
the K~dwarf, is more massive than the mass recipient, the WD,
but this is the most probable \citep{obrien01}.
In that case, the favorable scenario of a recent period
increase is mass loss at the level of 
$\sim 3.8 \times 10^{-7} M_{\sun} / yr$ \citep{hilditch01}.
Taking the masses of both components at the limits of the
\citet{obrien01} ranges we can also consider the case when
the donor is the less massive star. Such a situation would require
conservative mass transfer at a level of
$\sim 3.6 \times 10^{-6} M_{\sun} / yr$ 
to explain the recent period increase.
Both mass-loss rates appear to be large and unlikely for V471~Tau
as they would result in other detectable phenomena.
Moreover, both a period increase and a period decrease
have been observed for the system so the complete
picture would have to be even more complex.

The latest period change took place over some $\Delta E \simeq 2500$
cycles so the inferred time scale, $T$, was
$T = (d\ln P/dt)^{-1} \simeq 2 \times 10^6$ years.
This is a relatively short time scale for any thermal 
equilibrium adjustment in the K dwarf, but of course may
relate only to the outer layers of its envelope.

The standard deviation 
in the residuals of the second segment of 
$\sigma$ = 22.7~s (Figure~\ref{fig7}) is slightly 
larger than for any of the previous fits 
(14.9~s for the third-body model and 16.6~s for 
the apsidal motion model) but is still acceptable 
if superimposed upon possible short-timescale variations 
which are considered below.

\placefigure{fig7}

\subsubsection{Periodic residuals from eclipse timing models}
\label{residuals}

Every one of the $O-C$ models we calculated generates 
residuals with $\sigma$ larger than the accuracy of the
eclipse timings (typically a few seconds). We performed a 
search for periodicities in the residuals and found that 
regardless of the model used, there is evidence for a 10-year 
period in the timing residuals.

To investigate this further, we decided to employ a multi-harmonic 
analysis of variance (MAOV) with 2 harmonics, 
as described in \citet{alex96}. This method uses orthogonal 
polynomials to model the data and the analysis of 
variance statistics to evaluate the quality of the result. 
The MAOV periodogram obeys Fisher's probability distribution 
with $2N + 1$ and $K - 2N - 1$ degrees of freedom, 
where $N$ is the number of harmonics used and $K$ is the 
number of observations. The quantity $F(2N + 1, K - 2N - 1)$ 
shown in Figure~\ref{fig8} measures 
the ratio of powers of the periodic signal and 
residual, fully random noise.

The amplitude of the variations we find in the $O-C$ residuals 
is similar for all three models we adopted, at the 
level of 20--25~s and indeed indicates a typical underlying 
variation with a time scale of about 10 years.  The 
5.5-yr period found by \citet{ibanoglu05} -- which was 
connected with the ${\sim}5$-yr period in the mean 
brightness variations of the system -- is also present, 
but at a much lower significance level
(see Figure~\ref{fig8}).

\placefigure{fig8}

\subsection{Short-period oscillations}
\label{periodograms}

Fluctuations with a period of 555 s were discovered in 
soft X-ray flux from the V471~Tau system by the 
EXOSAT satellite \citep{jensen86}. In 1991, 131 hours 
of continuous U-band photometry of V471~Tau by the 
Whole Earth Telescope (WET) \citep{clemens92} resulted 
in the detection of three periods: 554.63, 561.59 
and 277.319 s.  The dominant 555-s variability 
(with its 277s harmonic) was attributed directly to
the accreting magnetic polar caps on the white dwarf component 
of the system, and the 562s signal to the
same radiation reprocessed by the K dwarf atmosphere.

To search for short-period variations in the MOST photometry, 
we first removed variations caused by the binary 
revolution and rotation of the spotted component. 
The data were ``rectified" by fitting the data with least-squares
low-order polynomials and then dividing by the 
fitted function. The eclipses and flare events (see 
Section \ref{flares} below), accounting for about 7\% of 
the total time series, were excluded from the fit,
resulting in a net duty-cycle of 59\%. The 
remaining 52,371 brightness measurements of the binary, 
as well as corresponding measurements of both 
reference stars, were used to calculate MAOV periodograms, 
as described above in Subsection~\ref{residuals}.

Analysis of the resulting periodogram revealed that none of 
the three WET periods is present in the MOST data, 
but their absence is easy to understand. While the white 
dwarf contribution to the total brightness of the system 
in the $U$ band is about 39\%, it is only 2.3\% in the 
broad MOST photometric bandpass which includes 
considerable red flux. Therefore, the relative amplitude of 
the variations in MOST photometry is expected to be 
about $17$ times smaller than in WET photometry.  The 
relative signal would be ${\sim}1.8 \times 10^{-4}$, 
which is slightly below our estimated, one-sigma detection limit of 
about $2 \times 10^{-4}$ in these data. This value was
calculated by folding the data with a period incommensurate 
with any of the V471 Tau variations and MOST orbital harmonics.
The noise estimation was also confirmed with the photometric data 
of both reference stars.

Thus, the non-detection of the white dwarf pulsations 
in the broad MOST passband is entirely predictable. We can 
conclude only that the pulse amplitude (and presumably 
the polar accretion rate) did not increase 
significantly since the WET campaign in 1991.

\subsection{Flare activity}
\label{flares}

Several flare-like events have been reported in  V471~Tau 
by \citet{rucinski81}, \citet{tunca93}, \citet{ibanoglu05} 
and others. \citet{young83} found that flares are most 
likely to occur when the brightness of the system is near 
its minimum, when the K dwarf was thought to have 
its most spotted hemisphere facing Earth.

In the MOST light curve, we identified seven events we 
would consider flare-like, although two of them were only 
partially recorded due to gaps in the data.
This is the first detection of white-light flares 
by the MOST satellite and probably the largest homogeneous
set of V471 Tau flare-like events observed so far.  The durations of 
these events varied from about 10 to over 35 minutes, 
but their shapes all share the same rapid rise and slower 
decay characteristic of flares seen in visible light.  
The candidate events are shown in Figure~\ref{fig9}.

\placefigure{fig9}

\placetable{tab4}

In contrast to \citet{young83}, we did not find any correlation 
of the flare events with the photometric wave minimum.
The flares occurred during phases of the lowest as well 
as the highest spottedness of the K dwarf, with no apparent 
concentration in phase.  The symbols at the bottom of 
Figure~\ref{fig4} mark the phases when the flares occurred.
Using luminosities of both components in the $V$ band given 
by \citet{obrien01}, we estimated a lower limit to the 
energy released during the whole duration of a typical 
flare observed during the MOST run at about $10^{34}$~erg 
(see Table~\ref{tab4}).  The energies of each of the seven 
flares we observed are comparable to the energy released by 
the flare reported by \citet{ibanoglu05} and are at the 
top of the range of energies released by all flare-like events 
reported for V471 Tau.
Because the activity cycle of V471~Tau still remains to
be characterized in terms of its period and intensity,
we cannot relate the observed incidence of flares to
the phase in this cycle. We note only that
all the observed flares share the shape, duration and energy
with those reported for typical RS~CVn systems.

The number of detected flare-like events corresponds 
to a total number of about 10 such events during the 10-day span 
of the MOST observations. Considering the limitations
of ground-based observations one would expect to be able to detect 
a maximum of 4 flare-like events during the same period of time.

\section{INTERPRETING THE SPECTRA}
\label{interpret-sp}

The typical S/N of the DDO spectra of V471~Tau is about 30.  
The contribution of the white dwarf component to the 
total light in the observed wavelength range is less than 1\%, 
so its contributions to the spectroscopic analyses 
described below are negligible. Our discussion of the
spectroscopic results is limited to the K dwarf in the system.

\subsection{Radial velocities}
\label{rv}

To derive the radial velocities (RV) of the K dwarf, we 
used the Broadening Function (BF) technique \citep{rucinski99}.
Spectra of four different K-type standard stars 
(HD~62509, HD~65583, HD~3765, HD~103095) were adopted as 
templates. The resulting broadening functions were fitted 
by a rotational line-broadening profile, with a linear 
limb-darkening coefficient of 0.52 (assumed to be typical 
for a K-type dwarf in the observed wavelength range), 
following \citet{vanhamme93}. The resulting RV measurements 
are listed in Table~\ref{tab1}.

We performed two independent least-squares fits to the 
radial velocities, assuming first a circular and then an eccentric
orbit, at a fixed orbital period as given 
by \citet{guinan01}, but with the time of conjunction taken from 
the MOST light curve.  The results of the fits and 
their residuals are plotted in Figure~\ref{fig10}.  The quality of 
both fits, evaluated by calculating the standard 
deviations of the residuals, is essentially identical for both types of 
orbits, with ${\sigma} \simeq 1.25$~km~s$^{-1}$. 
The fact that $\sigma$ is not reduced for a model with 
more free parameters suggests that the eccentric orbit 
solution is not necessary \citep{lucy71}, although obviously 
this is not a proof for perfect circularity of the V471~Tau orbit.

\placefigure{fig10}

\placetable{tab5}

All our orbital model parameters (Table~\ref{tab5}) 
agree very well with those obtained recently by 
\citet{hussain06}, but they deviate slightly from those 
obtained previously with the same DDO 1.88-m 
telescope by \citet{bois88}.  The amplitude we find is 
larger by about $1.5 - 2$~km~s$^{-1}$, and the center-of-mass 
radial velocity is about 2~km~s$^{-1}$ smaller.

\subsection{Projected rotation velocity}
\label{rotation}

A bonus of the BF analysis is the availability of the 
projected rotation profile of the star onto radial velocity space
(Figure~\ref{fig11}). This shape 
can be interpreted through a solid-body rotation to 
estimate the projected equatorial velocity 
$V_{K} \sin i$. In the BF determination, we used HD~3765 as a 
standard star because its spectral type, K2V, is 
identical to that of the V471~Tau K dwarf. An average of the 
projected rotational velocities for all spectra is 
$V_K \sin i = 91.9 \pm 2.5$~km~s$^{-1}$. The value is corrected 
for the broadening introduced by the method, the 
magnitude of which can be estimated by applying the BF 
method to the template itself.  The result is consistent 
with previous estimates made by \citet{ramseyer95} and 
\citet{hussain06} ($91 \pm 4$ and $91 \pm 2$ km~s$^{-1}$, 
respectively) and all are consistent with synchronous 
rotation of the K dwarf in V471~Tau.

\placefigure{fig11}

\subsection{$H\alpha$ emission}
\label{halpha}

The $H\alpha$ line was detected in emission in V471~Tau 
by \citet{lanning76}. Subsequent detailed studies 
\citep{young88, bois91, rottler02} revealed orbital 
phase-dependence of the emission as well as long-term 
changes of its equivalent width.

We extracted the $H\alpha$ emission from the absorption 
profiles of our spectra by again using the standard 
star HD~3765 as a template.  HD~3765 has the same spectral 
type as the V471~Tau K dwarf and rotates very slowly 
at $V \sin i \simeq 1$~km~s$^{-1}$ \citep{soderblom85}. 
We convolved the standard spectrum with the 
rotational profile calculated for $V_K \sin i = 92$~km~s$^{-1}$ 
(our value for V471~Tau) and fitted the resulting 
modified spectrum to each of our V471~Tau spectra in 
two wavelength ranges: $6540 - 6555$~\AA\ and $6570
- 6585$~\AA\ (see Figures~\ref{fig3} and \ref{fig12}).  
Subsequently, we used the net $H\alpha$ 
emission to derive the radial velocities and equivalent 
widths of the emission line (Table~\ref{tab1}).
The extracted $H\alpha$ profiles were symmetrical
thus allowing us to use a Gaussian fit for measuring RV
and numerical integration for equivalent widths.
The radial velocity of the $H\alpha$ emission 
(Figure~\ref{fig13}) follows the K dwarf orbital 
variations, but with a smaller amplitude of about 
120~km~s$^{-1}$, as estimated from a sinusoidal fit.  Such 
behavior was observed during 1975 -- 1984 by 
\citet{bois91}, but with a still much smaller amplitude of $\sim 75$~
km~s$^{-1}$. We observe that the $H\alpha$ equivalent 
width changes symmetrically with respect to its 
maximum at orbital phase 0.5 (Figure~\ref{fig13}), 
in a very similar way to what was reported by \citet{bois91}. 
The amplitude of the equivalent width variability in 
our data is about 1.2~\AA\, with 
the maximum emission of about $-0.5$~\AA\ at phase 0.5.

\placefigure{fig12}

\placefigure{fig13}

Long-term changes of $H\alpha$ emission were detected 
by \citet{bois91}, who showed that the emission
strength diminished between 1975 and 1983 and then grew 
rapidly in 1984. More recent observations by 
\citet{rottler02} have shown that since 1985, the emission 
was dropping again, until it finally vanished in 
1992.  This suggests that the long-term variation in 
$H\alpha$ emission strength may be periodic, with a 
period of roughly 9~years.  Our measurements show 
that in December 2005, the emission strength was 
comparable to its average values in the past. This is 
consistent with a 9-year periodicity, since our 
DDO spectra were obtained about 2~years after the latest 
expected emission maximum in such a cycle.

\section{SUMMARY}
\label{summary}

The nearly continuous MOST spacebased photometry of 
V471~Tau covering 10 days in December 2005, 
combined with partly simultaneous DDO groundbased 
spectroscopy, monitored a fairly quiescent stage 
in the activity of the K dwarf in this close binary system.  
This is apparent in the light curve which deviates
relatively little from the model 
and almost does not change during the whole observing run.
Even during such a stable time, seven candidate 
flare events were observed in 10 days, whose estimated 
energies would be among the highest ever seen  
in V471~Tau.  There is no correlation between the times 
of the flares and orbital phase.

The main features of the orbital phase diagram of the 
MOST photometry are well reproduced by our eclipsing 
binary light curve synthesis model.  The largest systematic 
deviation in the double-wave light curve is only 
about $0.02 - 0.03$ mag and is consistent with spots 
on the K dwarf which is expected to rotate 
synchronously with the orbit.  The amount of spottedness 
on the star did not change much during the MOST 
observing run -- by no more than about 1\%. This supports the 
claim that the K dwarf was observed close to a 
minimum in its activity cycle. A half-orbital-period modulation 
of the radial velocity residuals was reported 
earlier by \citet{hussain06} and interpreted as an asymmetry in 
spot distribution on the K star's surface. We 
see no such residuals in our radial velocity measurements.  
We note that the residuals seen by 
\citet{hussain06}, the radial velocity curve we obtain, and 
the $O-C$ variations in eclipse times observed over 
the past 35 years, could all be interpreted as a small 
non-zero eccentricity of the orbit of V471~Tau.

Because of the broad bandpass of the MOST photometry 
with substantial flux in the red, and the red 
wavelength range of the DDO spectra,  the white dwarf 
contributes only about 2\% and 1\% of the total intensity of the 
system, respectively.  We were therefore unable to 
constrain the properties of the hot white dwarf in 
the system or confirm the oscillation frequencies detected by 
WET  \citep{clemens92}, since the relative 
amplitudes in the custom-filter, broadband MOST photometry would 
be about 17 times smaller than in the WET 
$U$-band photometry.  The positive aspect of this is that 
our estimates of the K dwarf properties from MOST 
photometry and DDO spectroscopy are not contaminated by 
the white dwarf, but we can use the timing of 
the white dwarf eclipses to investigate aspects of the 
orbit of the V471~Tau system.

\begin{figure}    
\epsscale{1.0}
\plotone{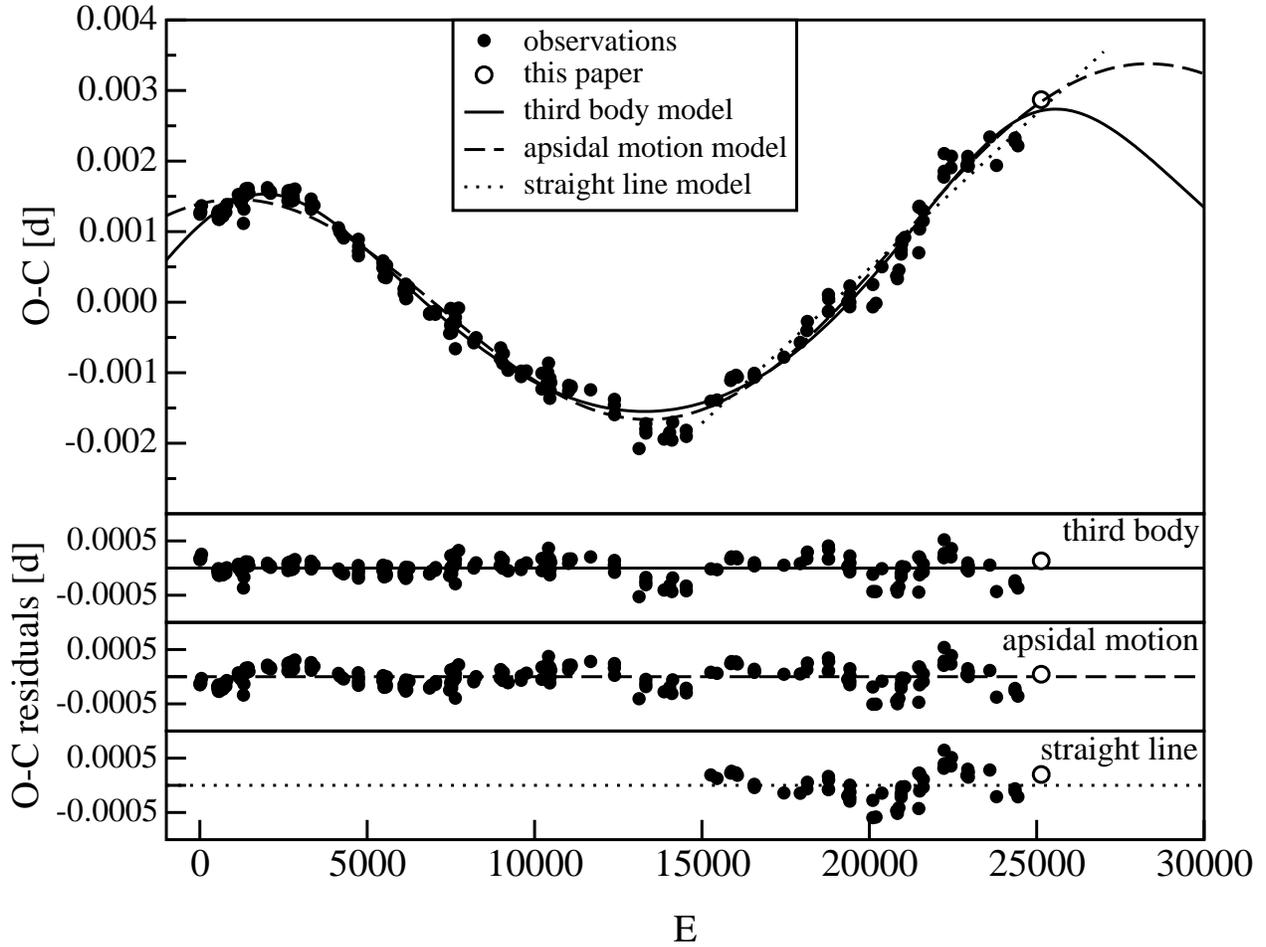}
\caption{Comparison of the third body, apsidal motion and 
straight line model fits to the available eclipse time 
observations of V471 Tau. \label{fig7}
}
\end{figure}

\begin{figure}    
\epsscale{1.0}
\plotone{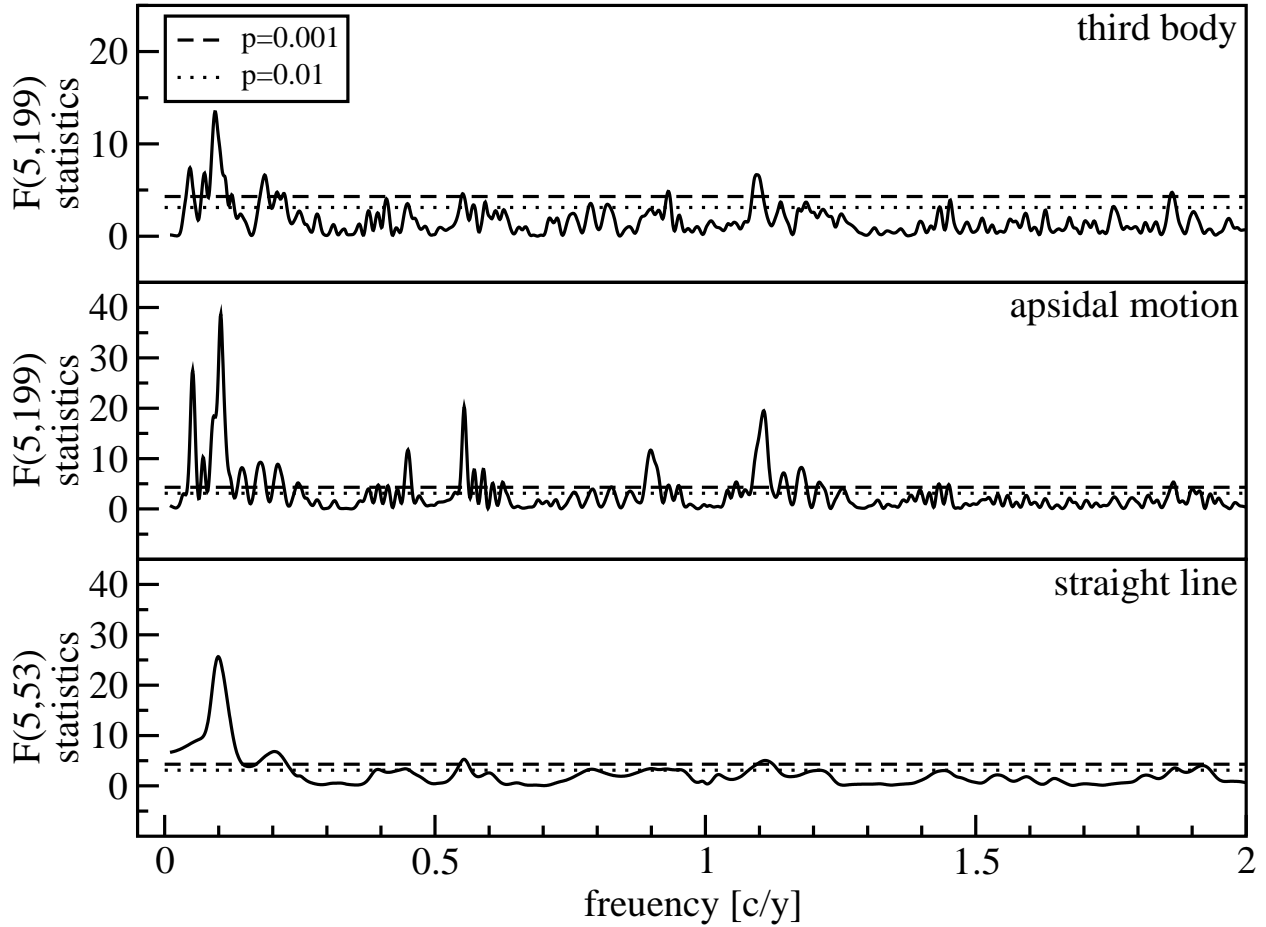}
\caption{Multiharmonic analysis of variance
periodograms \citep{alex96}, with frequencies up to 2 cycles 
per year, for different model residuals, as discussed 
in Subsection~\ref{timing}. Horizontal lines show the
levels of 0.001 (dashed line) and 0.01 (dotted line)
probability of false detection.
The most significant peaks appear around
the same frequency $\sim 0.1$ c/y for all models 
(the top two periodograms also show its alias at $\sim 0.05$~c/y).
Note that the 5.5~yr period (0.18~c/y) found by \citet{ibanoglu05}
appears to be also present. \label{fig8}
}
\end{figure}

\begin{figure}    
\epsscale{1.0}
\plotone{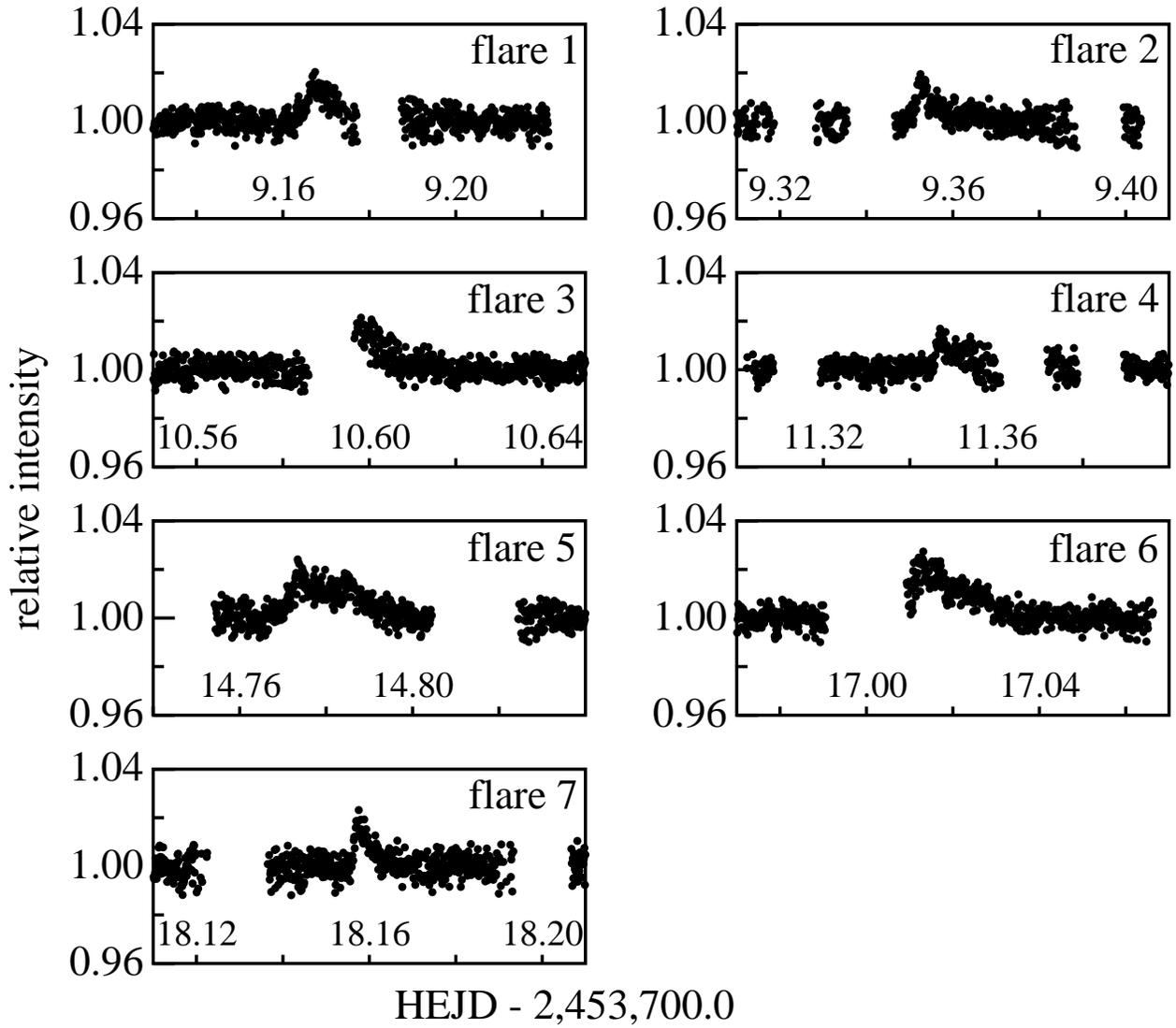}
\caption{The seven flare-like events on V471 Tau
which were detected during MOST observation period. \label{fig9}
} 
\end{figure}

\begin{figure}    
\epsscale{0.85}
\plotone{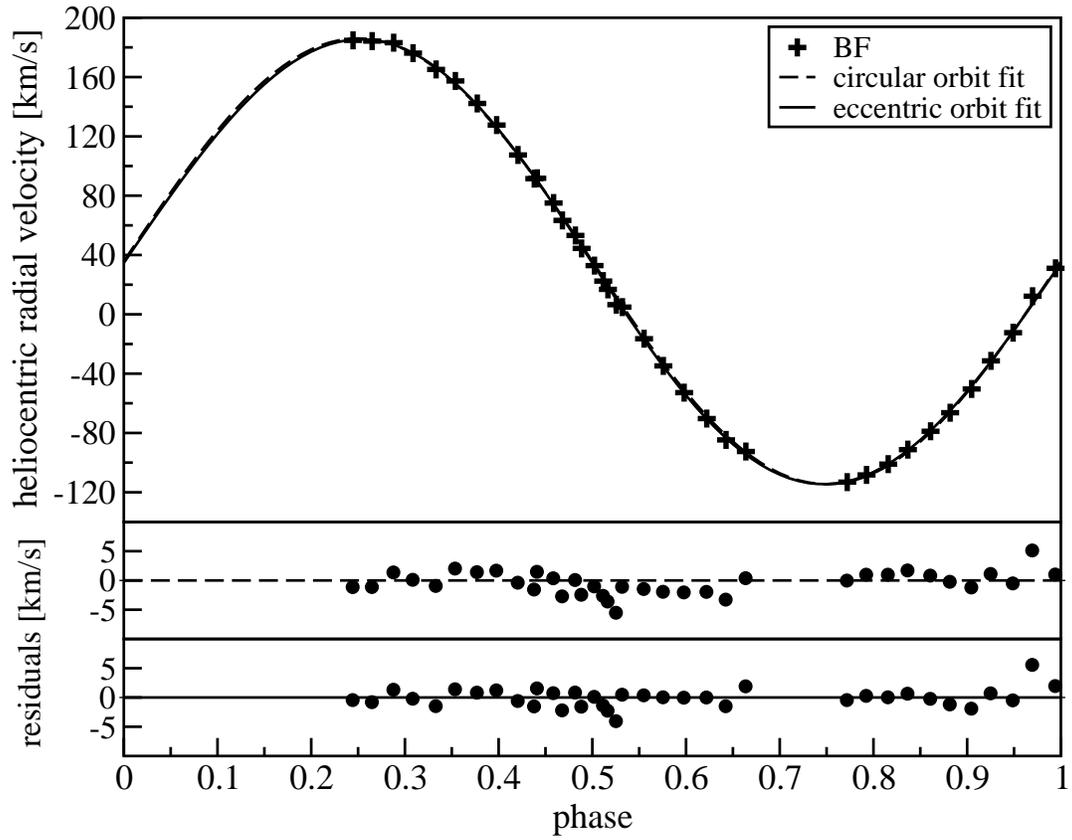}
\caption{The radial velocity curve for the K dwarf component 
of V471~Tau binary. The bottom panels show residuals for the
circular and elliptical models, respectively. \label{fig10}
}
\end{figure}

\begin{figure}    
\epsscale{0.85}
\plotone{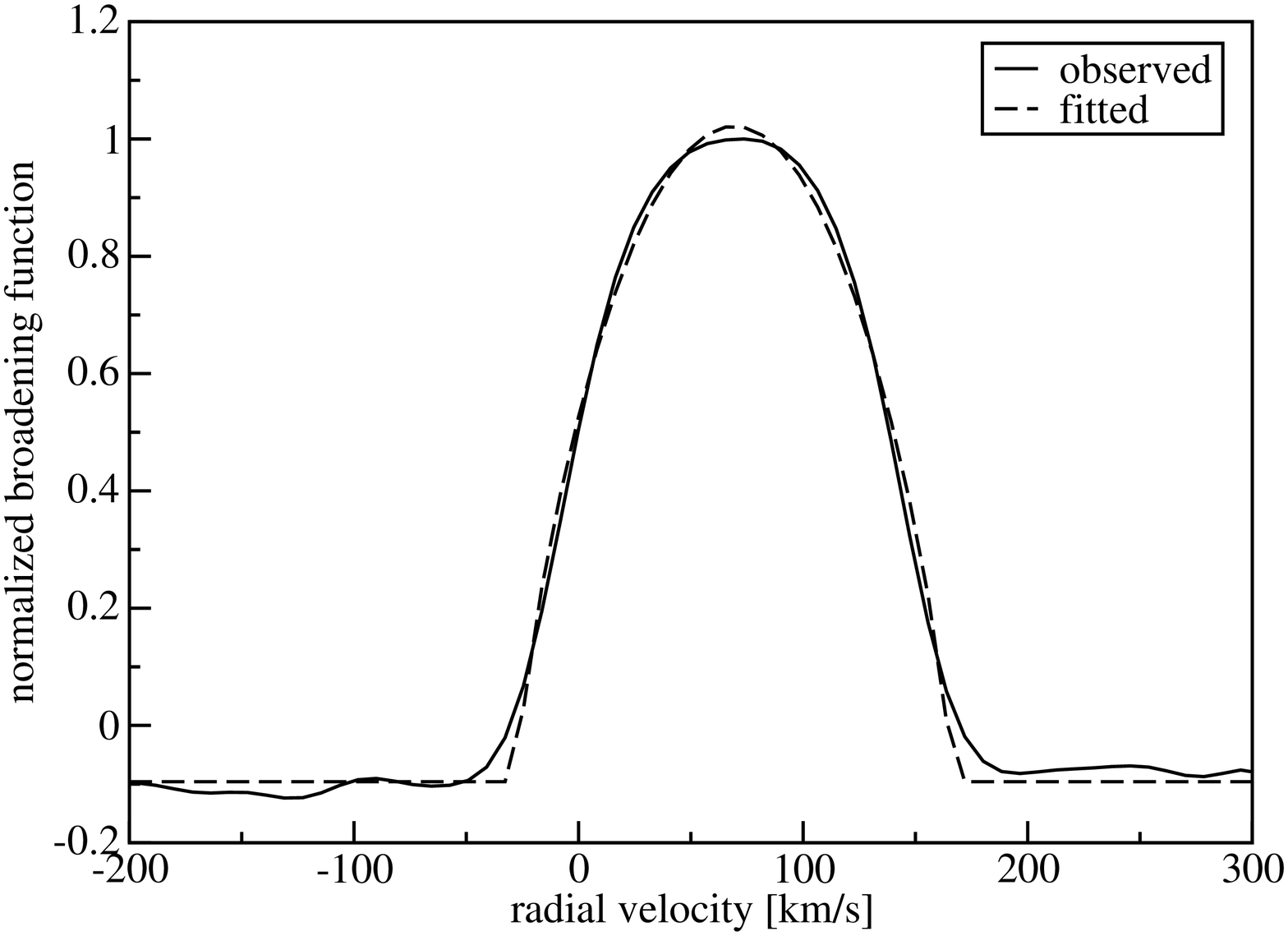}
\caption{The average broadening function of V471~Tau 
spectra derived with
the standard velocity star HD~3765 of the same spectral
type (solid line). This BF profile was fitted by the rotational
broadening profile to estimate the projected rotation
velocity of the K dwarf component (dashed line). 
\label{fig11}
}
\end{figure}

\begin{figure}    
\epsscale{0.85}
\plotone{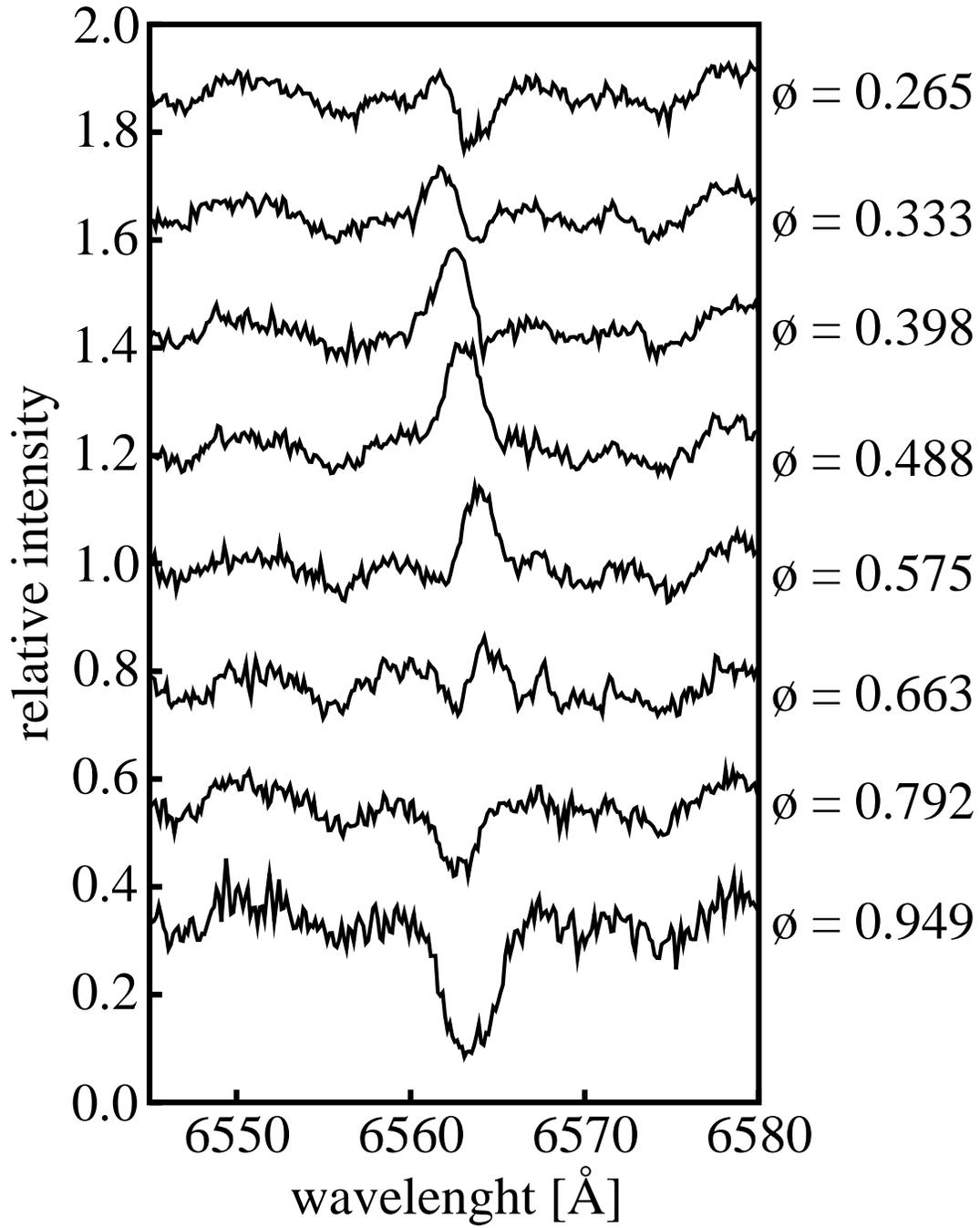}
\caption{A collection of representative spectra of V471 Tau
taken at different phases.
The variable strength and shifts in position of the $H\alpha$
emission are clearly visible. \label{fig12}
}
\end{figure}

\begin{figure}    
\epsscale{0.85}
\plotone{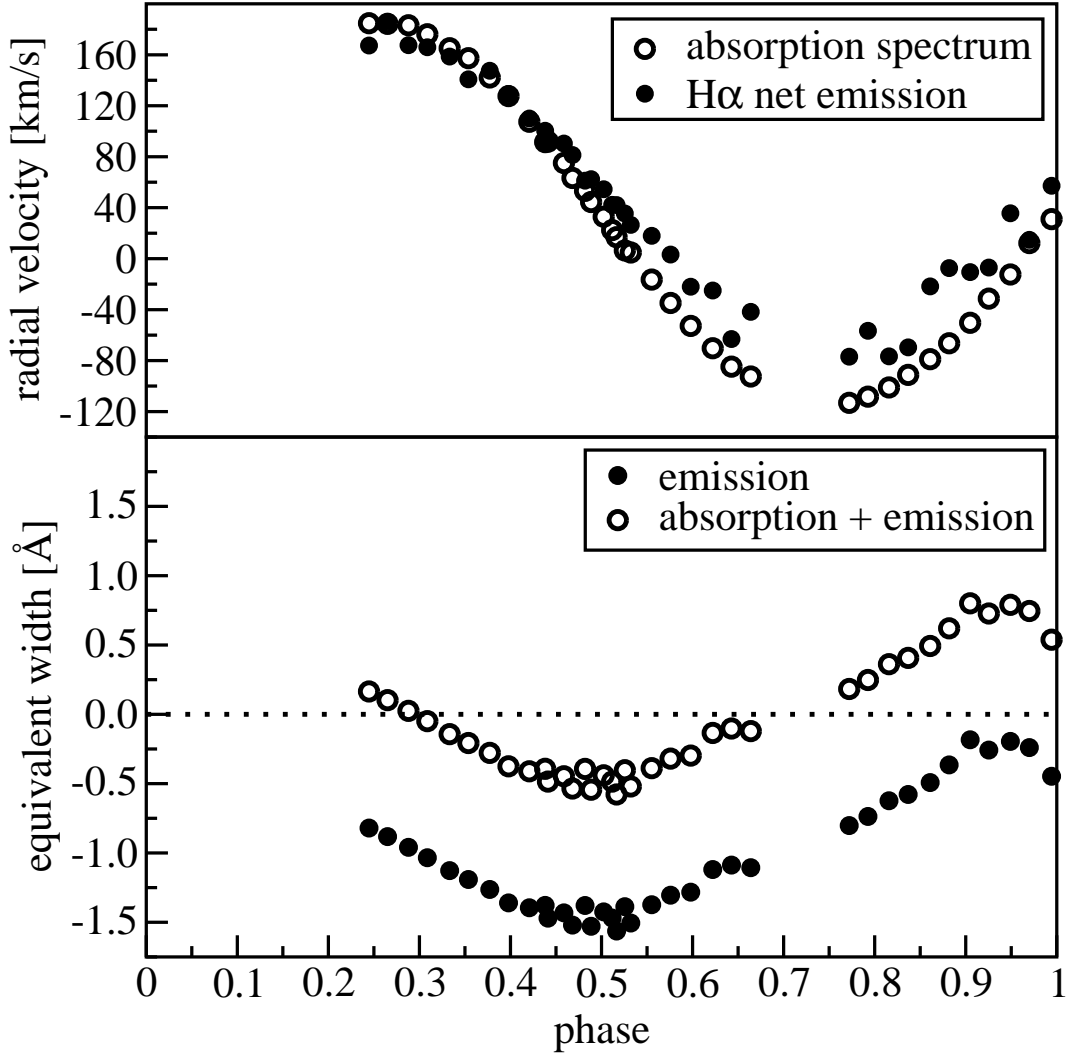}
\caption{Variations of the $H\alpha$ emission line. 
Top: The radial velocity changes of the net emission
with phase, compared with those of the K-dwarf itself (the
absorption spectrum). Note the reduced amplitude of about 120~km~s$^{-1}$.
Bottom: Changes of the $H\alpha$ line equivalent
width. The strongest emission is visible when the K-dwarf
component is seen in the upper conjunction (when the face illuminated
by the WD is directed to the observer around the 
orbital phase of 0.5). The emission is practically  
undetectable during the opposite phases. \label{fig13}
}
\end{figure}

Changes in the $O-C$ values of the times of eclipse of 
the white dwarf can, however, be explained by at 
least three entirely different models: (1)~There could have been at 
least two abrupt period changes in the orbit 
of the system in the last 35 years, although there is no 
obvious mechanism for this. (2)~There could be 
apsidal motion due to a slightly eccentric orbit.  
(3)~The V471~Tau system might be a trinary, with a third 
low-mass companion in a long-period orbit.  The last two 
periodic phenomena both predict that the $O-C$ eclipse timing
deviations must drop in the future (see Figure \ref{fig7}). 
The small eccentricity which could 
explain the $O-C$ diagram is also in agreement with the 
formal solution of the radial velocity curve of the K 
dwarf from our high-quality DDO spectra, but its value  
is currently below the direct spectroscopic detection threshold.
Future accurate eclipse timing observations, such as performed
by the MOST satellite, are desired as they may resolve
the dilemma between those three models.

The $O-C$ residuals do show a convincing residual periodic 
variation with a period of about 10 years,
regardless of the model used to explain the longer-term changes. 
This variation may be due to an activity 
cycle in the K dwarf, but this is a highly speculative 
explanation. We note that the $H\alpha$ emission 
appears to change in intensity in a characteristic time 
scale of about 9 years, perhaps coincident with the 
periodicity in eclipse time variations at the frequency 
resolution of the entire data sample at hand.
The 10-year period in the $O-C$ residuals may also be
related with the 5.5-year period in the system mean brightness variations
found by \citet{ibanoglu05} as its multiple. Nevertheless we think
that both periods are too uncertain to firmly connect them
at this stage of the study of V471 Tau.

The DDO spectra yield a new radial velocity curve for the 
orbit of the K dwarf, and an improved determination 
of the projected rotation of the star, $V \sin i = 92$ km~s$^{-1}$ 
based on high-quality BF (broadening function) 
profiles.  The spectra also enabled us to measure the 
$H\alpha$ emission velocities and changes in its 
equivalent width. The $H\alpha$ emission of V471~Tau 
showed the same orbital phase dependence as 
observed before by \citet{bois91} and \citet{rottler02} 
with maximum emission at phase ${\sim}0.5$. The 
observed amplitude of equivalent width variations of 
about 1.2~\AA\ was average for the system and 
consistent with the 9-year period noted by previous 
investigators. Unfortunately, the 13-year gap between 
the most recent published $H\alpha$ emission observations 
of V471~Tau and our new DDO observations 
does not allow us to reliably verify the periodic character 
of the mean emission strength variation.

A new feature of the $H\alpha$ emission revealed by 
our observations was its much larger amplitude of 
radial velocity variation (120~km~s$^{-1}$) 
compared to that reported by earlier observers
(75~km~s$^{-1}$ by \citet{bois91}).  
This suggests that the source of the emission was less
concentrated around the sub-white-dwarf point on the 
K star as had been seen in the previous data.

\acknowledgements

The research of SMR, JMM, DBG, AFJM, DS and GAHW 
was supported by grants from NSERC 
(Natural Sciences and Engineering Council) Canada.
WWW is supported by the Aeronautics and Space Agency of FFG
and the Austrian Science Fund (FWF) P17580.
RK is supported by the Canadian Space Agency 
through a contract with UBC.
AFJM is supported from FQRNT (Quebec).
KK appreciates the hospitality 
and support of the local staff during his stay 
at DDO. Special thanks are due to the DDO Telescope Operators, 
Heide DeBond and Jim Thomson, for 
help with the spectroscopic observations, and to MOST 
Satellite Operators, Alex Beattie, Jamie Wells 
and Ron Wessels.

\clearpage


\begin{deluxetable}{ccccccc}

\tabletypesize{\scriptsize}
\tablewidth{0pt}

\tablecaption{Spectroscopic observations.
\label{tab1}}
\tablehead{
\colhead{No.} &
\colhead{$HJED - 2,453,700$} &
\colhead{photometric} &
\colhead{S/N} &
\colhead{$V_{rad}$} &
\colhead{$H\alpha$ EW} &
\colhead{$v_{rad}$ of $H\alpha$} \\
\colhead{} &
\colhead{} &
\colhead{phase} &
\colhead{} &
\colhead{(km~s$^{-1}$)} &
\colhead{(\AA)} &
\colhead{net emission (km~s$^{-1}$)}
}
\startdata
 1  & 17.58275 & 0.5162 & 30 &   16.8 & -0.578 &  42 \\
 2  & 17.71561 & 0.7717 & 30 & -113.2 &  0.183 & -77 \\
 3  & 17.72632 & 0.7922 & 30 & -108.3 &  0.249 & -57 \\
 4  & 17.73841 & 0.8154 & 30 & -101.1 &  0.361 & -77 \\
 5  & 17.74935 & 0.8364 & 30 &  -91.2 &  0.407 & -70 \\
 6  & 17.76196 & 0.8606 & 30 &  -78.8 &  0.493 & -22 \\
 7  & 17.77283 & 0.8815 & 30 &  -66.3 &  0.620 &  -7 \\
 8  & 17.78482 & 0.9045 & 12 &  -50.3 &  0.801 & -11 \\
 9  & 17.79554 & 0.9250 & 30 &  -31.3 &  0.727 &  -7 \\
 10 & 17.80788 & 0.9487 & 30 &  -12.4 &  0.789 &  36 \\
 11 & 17.81871 & 0.9695 &  8 &   12.2 &  0.745 &  15 \\
 12 & 17.83145 & 0.9939 &  8 &   31.0 &  0.538 &  57 \\
 13 & 18.48312 & 0.2443 & 20 &  184.8 &  0.165 & 167 \\
 14 & 18.49374 & 0.2647 & 30 &  184.2 &  0.103 & 186 \\
 15 & 18.50572 & 0.2877 & 30 &  183.2 &  0.025 & 168 \\
 16 & 18.51648 & 0.3083 & 30 &  176.1 & -0.050 & 166 \\
 17 & 18.52928 & 0.3329 & 30 &  165.2 & -0.143 & 159 \\
 18 & 18.54004 & 0.3535 & 30 &  157.4 & -0.207 & 141 \\
 19 & 18.55221 & 0.3768 & 30 &  142.2 & -0.278 & 147 \\
 20 & 18.56299 & 0.3975 & 30 &  127.6 & -0.375 & 129 \\
 21 & 18.57489 & 0.4203 & 30 &  107.5 & -0.411 & 110 \\
 22 & 18.58558 & 0.4409 & 30 &   91.8 & -0.485 &  92 \\
 23 & 18.59955 & 0.4677 & 30 &   63.3 & -0.535 &  81 \\
 24 & 18.61024 & 0.4882 & 30 &   44.4 & -0.544 &  62 \\
 25 & 18.62229 & 0.5113 & 30 &   22.4 & -0.485 &  42 \\
 26 & 18.63295 & 0.5318 & 30 &    4.8 & -0.520 &  26 \\
 27 & 18.64494 & 0.5548 & 30 &  -16.4 & -0.388 &  18 \\
 28 & 18.65570 & 0.5754 & 30 &  -34.7 & -0.319 &   3 \\
 29 & 18.66658 & 0.5976 & 15 &  -52.8 & -0.298 & -22 \\
 30 & 18.67986 & 0.6218 & 30 &  -70.3 & -0.135 & -25 \\
 31 & 18.69057 & 0.6423 & 30 &  -84.7 & -0.102 & -63 \\
 32 & 18.70082 & 0.6635 & 30 &  -92.5 & -0.121 & -42 \\
 33 & 23.79578 & 0.4377 & 30 &   91.5 & -0.393 & 100 \\
 34 & 23.80649 & 0.4583 & 30 &   75.0 & -0.447 &  90 \\
 35 & 23.81862 & 0.4816 & 10 &   53.2 & -0.394 &  61 \\
 36 & 23.82925 & 0.5020 & 10 &   32.9 & -0.440 &  55 \\
 37 & 23.84138 & 0.5252 & 20 &    6.5 & -0.403 &  35 \\
\enddata
\end{deluxetable}


\begin{deluxetable}{ll}

\tabletypesize{\scriptsize}
\tablewidth{0pt}

\tablecaption{The best fit parameters for a third body model.
\label{tab2}}
\tablehead{
\colhead{parameter} &
\colhead{value}
}
\startdata
 $T_{0}$ (HJED)                      & $2440610.06446\: \pm \:0.00008$    \\ 
 $P_{0}$ (d)                         & $0.521183449\: \pm \:0.000000008$  \\ %
 $P_{3}$ (yr)                        & $33.7\: \pm \:0.9$               \\ 
 $a_{12} \sin(i_{3})$ (AU)           & $0.32\: \pm \:0.02$              \\ 
 semiamplitude (sec)                   & $159\: \pm \:6$                  \\ 
 $e$                                 & $0.28\: \pm \:0.09$              \\ 
 $\omega_{3}$ (degr)                & $79\: \pm \:10$                  \\ 
 $f(M_{3})$ ($M_{\sun}$)             & $(2.9 \pm 0.3) \cdot 10^{-5}$    \\ 
\enddata
\tablecomments{The errors were estimated
using the bootstrap sampling method for the 98\% confidence level;
this is why they appear to be large when compared to other fits,
for example by \citet{ibanoglu05}.}

\end{deluxetable}


\begin{deluxetable}{ll}

\tabletypesize{\scriptsize}
\tablewidth{0pt}

\tablecaption{The best fit parameters for the orbital apsidal motion.
\label{tab3}}
\tablehead{
\colhead{parameter} &
\colhead{value}
}
\startdata
 $P$ (day)                             & $38.884\: \pm \:0.007$     \\
 semiamplitude (sec)                   & $173\: \pm \:9$      \\
 $e$                                 & $0.0121\: \pm \:0.0006$    \\
 $\omega_{0}$ (degr)                & $174\: \pm \:3$     \\
 $\omega_{1}$ (degr/day)              & $0.025348\: \pm \:0.000005$  \\
\enddata
\end{deluxetable}


\begin{deluxetable}{cccccc}

\tabletypesize{\scriptsize}
\tablewidth{0pt}

\tablecaption{A list of flare-like events observed by MOST.
\label{tab4}}
\tablehead{
\colhead{nr} &
\colhead{start time (HJED)} &
\colhead{phase} &
\colhead{duration (min)} &
\colhead{$\Delta I_{max}$} &
\colhead{$E_{min}$ (erg)}
}
\startdata
1   &  2453709.164  &  0.379  &   19    &  0.014    &  $9.8\cdot 10^{33}$  \\
2   &  2453709.351  &  0.738  &   12    &  0.014    &  $3.2\cdot 10^{33}$  \\
3*  &  2453710.59   &  0.12   &  $>16$  & $>0.016$  &  $>1.8\cdot 10^{34}$ \\
4   &  2453711.345  &  0.564  &   13    &  0.011    &  $1.0\cdot 10^{34}$  \\
5   &  2453714.770  &  0.135  &   36    &  0.018    &  $2.9\cdot 10^{34}$  \\
6*  &  2453717.01   &  0.43   &  $>33$  &  0.018    &  $>4.4\cdot 10^{34}$ \\
7   &  2453718.156  &  0.632  &   10    &  0.017    &  $1.9\cdot 10^{34}$  \\
\enddata
\tablecomments{$E_{min}$ designates the lower limit for total energy
released in the $V$ band. The flares marked with a star were 
observed only partially.}
\end{deluxetable}


\begin{deluxetable}{ccc}

\tabletypesize{\scriptsize}
\tablewidth{0pt}

\tablecaption{Parameters of spectroscopic orbits.
\label{tab5}}
\tablehead{
\colhead{parameter} &
\colhead{circular orbit} &
\colhead{eccentric orbit}
}
\startdata

$K_{K}$ (km~s$^{-1}$)          & $150.5 \pm 0.4$   & $150.2 \pm 0.5$ \\
$a \sin i$ ($R_{\sun}$)         & $1.550 \pm 0.004$ & $1.547 \pm 0.005$ \\
$e$                            & --                & $0.012 \pm 0.003$ \\
$\omega_{1}$  (degr)            & --                & $75 \pm 18$ \\
$V_0$ (km~s$^{-1}$)            & $35.7 \pm 0.3$    & $35.2 \pm 0.3$ \\
$\sigma$ (km~s$^{-1}$) & $1.25$            & $1.25$      \\

\enddata
\end{deluxetable}


\begin{deluxetable}{cccccccc}

\tabletypesize{\scriptsize}
\tablewidth{0pt}

\tablecaption{Parameters of hypothetical third body in the V471 Tau system.
\label{tab6}}
\tablehead{
\colhead{$i_{3}$ (degr)} &
\colhead{$M_{3}$ ($M_{\sun}$)} &
\colhead{$T_{eff}$ (K)} &
\colhead{$log L/L_{\sun}$} &
\colhead{$m_{V}$} &
\colhead{$m_{K}$} &
\colhead{$d_{max}$ (mas)} &
\colhead{$T_{max}$ (year)}
}
\startdata

$85$ &  $0.045$ &  $1540$ &  $-4.3$ &  $29.6$ &  $15.4$ &  $420$ &  $2014.1$ \\
$60$ &  $0.052$ &  $1730$ &  $-4.1$ &  $26.8$ &  $15.0$ &  $490$ &  $2015.0$ \\
$45$ &  $0.064$ &  $2060$ &  $-3.8$ &  $24.0$ &  $14.4$ &  $610$ &  $2016.1$ \\
$30$ &  $0.090$ &  $2660$ &  $-3.2$ &  $19.2$ &  $13.3$ &  $910$ &  $2019.0$ \\

\enddata
\tablecomments{The physical parameters of the third body are based on nongray dusty models
of \citet{chabrier02}, assuming the age of the system of 625 Myr and the dystance of 46.8 pc.
$d_{max}$ designates maximum apparent separation between the V471 Tau binary
and the third component. $T_{max}$ is the time of the nearest maximum separation.}
\end{deluxetable}


\clearpage

\begin{thebibliography}{}




\bibitem[Bastian(2000)]{bastian2000}
    Bastian, U.
    2000, Inf.\ Bull.\ Var.\ Stars, 4822



\bibitem[Bois et al.(1988)]{bois88}
    Bois, B., Lanning, H. H., Mochnacki, S. W.
    1988, \aj, 96, 157

\bibitem[Bois et al.(1991)]{bois91}
    Bois, B., Lanning, H. H., Mochnacki, S. W.
    1991, \aj, 102, 6




\bibitem[Chabrier et al.(2000)]{chabrier02}
    Chabrier, G., Baraffe, I., Allard, F., Hauschildt, P.
    2000, \apj, 542, 464

\bibitem[Clemens et al.(1992)]{clemens92}
    Clemens, J. C., Nather, R. E., Winget, D. E., Robinson, E. L., 
    Wood, M. A., Claver, C. F.,     Provencal, J., Kleinman, S. J., 
    Bradley, P. A., Frueh, M. L., Grauer, A. D., Hine, B. P.,
    Fontaine, G., Achilleos, N., Wickramasinghe, D. T., 
    Marar, T. M. K., Seetha, S.,  Ashoka, B. N., O'Donoghue, D., 
    Warner, B., Kurtz, D. W., Martinez, P.,
    Vauclair, G., Chevreton, M., Barstow, M. A., Kanaan, A., Kepler, S. O.,
    Augusteijn, T., van Paradijs, J., Hansen, C. J.
    1992, \apj, 391, 773



\bibitem[Guinan \& Ribas(2001)]{guinan01}
    Guinan, E. F., Ribas, I.
    2001, \apj, 546, L43




\bibitem[Herczeg(1975)]{herczeg75}
    Herczeg, T. J.
    1975, Inf.\ Bull.\ Var.\ Stars, No. 1076

\bibitem[Hilditch(2001)]{hilditch01}
    Hilditch, R. W.
    2001, ``An Introduction to Close Binary Stars'', 
    Cambridge University Press, Cambridge

\bibitem[Hussain et al.(2006)]{hussain06}
    Hussain, G. A. J., Allende Prieto, C., Saar, S. H., Still, M.
    2006, \mnras, 367, 1699

\bibitem[Ibanoglu(1978)]{ibanoglu78}
    Ibanoglu, C.
    1978, \apss, 57, 219

\bibitem[Ibanoglu(1989)]{ibanoglu89}
    Ibanoglu, C.
    1989, \apss, 161, 221

\bibitem[Ibanoglu et al.(1994)]{ibanoglu94}
    Ibanoglu, C., Keskin, V., Akan, M. C., Evren, S., Tunca, Z.
    1994, \aap, 281, 811

\bibitem[Ibanoglu et al.(2005)]{ibanoglu05}
    Ibanoglu, C., Evren, S., Tas, G., Cakirli, O.
    2005, \mnras, 360, 1077

\bibitem[Jensen et al.(1986)]{jensen86}
    Jensen, K. A., Swank, J. H., Petre, R., Guinan, E. F.,
    Sion, E. M., Shipman, H. L.
    1986, \apj, 309, L27

\bibitem[Lanning \& Etzel(1976)]{lanning76}
    Lanning, H. H., Etzel, P. B.
    1976, Inf.\ Bull.\ Var.\ Stars, No. 1147




\bibitem[Lucy \& Sweeney(1971)]{lucy71}
    Lucy, L.~B., Sweeney, M.~A.
    1971, \aj, 76, 544

\bibitem[Matthews et al.(2004)]{matthews2004}
    Matthews, J. M., Kuschnig, R., Guenther, D. B.,
    Walker, G. A. H., Moffat, A. F. J., Rucinski, S. M.,
    Sasselov, D., Weiss, W. W.
    2004, Nature, 430, 921

\bibitem[Moffat (1969)]{moffat69}
    Moffat, A. F. J.,
    1969, \aap, 3, 455




\bibitem[O'Brien et al.(2001)]{obrien01}
    O'Brien, M. S., Bond, H. E., Sion, E. M.
    2001, \apj, 563, 971

\bibitem[Pr\v{s}a \& Zwitter(2005)]{prsa05}
    Pr\v{s}a, A., Zwitter, T.
    2005, \apj, 628, 426

\bibitem[Ramseyer et al.(1995)]{ramseyer95}
    Ramseyer, T. F., Hatzes, A. P., Jablonski, F.
    1995, \aj, 110, 3


\bibitem[Rowe et al.(2006)]{rowe06}
    Rowe, J.~F., Matthews, J.~M., Seager, S., Kuschnig, R.,
    Guenther, D.~B., Moffat, A.~F.~J., Rucinski, S.~M., Sasselov, D.,
    Walker, G.~A.~H., Weiss, W.~W.
    2006, \apj, 646, 1241

\bibitem[Rottler et al.(2002)]{rottler02}
    Rottler, L., Batalha, C., Young, A., Vogt, S.
    2002, \aap, 392, 535

\bibitem[Rucinski(1981)]{rucinski81}
    Rucinski, S.~M.
    1981, \actaa, 31, 37

\bibitem[Rucinski(1999)]{rucinski99}
    Rucinski, S.~M.
    1999, in ASP Conf. Ser. 185, Precise Stellar Radial Velocities,
    ed. J. B. Hearnshaw \& C. D. Scarfe (San Francisco: ASP), 82


\bibitem[Schwarzenberg-Czerny(1996)]{alex96}
    Schwarzenberg-Czerny, A.
    1996, \apjl, 406, 107

\bibitem[Skillman \& Patterson(1988)]{skillman88}
    Skillman, D. R., Patterson, J.
    1988, \aj, 96, 976

\bibitem[Soderblom(1985)]{soderblom85}
    Soderblom, D. R.
    1985, \aj, 90, 10

\bibitem[Still \& Hussain(2003)]{still03}
    Still, M., Hussain, G.
    2003, \apj, 597, 1059

\bibitem[Todoran(1972)]{todoran72}
    Todoran, I.
    1972, \apss, 15, 229

\bibitem[Tunca et al.(1993)]{tunca93}
    Tunca, Z., Keskin, V., Evren, S., Ibanoglu, C., Akan, M. C.
    1993, \apss, 204, 297


\bibitem[van Hamme(1993)]{vanhamme93}
    van Hamme, W.
    1993, \aj, 106, 2096

\bibitem[Walker et al.(2003)]{walker2003}
    Walker, G., Matthews, J., Kuschnig, R., Johnson, R.,
    Rucinski, S., Pazder, J., Burley, G., Walker, A.,
    Skaret, K., Zee, R., Grocott, S., Carroll, K.,
    Sinclair, P., Sturgeon, D., Harron, J.
    2003, \pasp, 115, 1023


\bibitem[Werner \& Rauch(1997)]{werner97}
    Werner, K., Rauch, T.
    1997, \aap, 324, L25



\bibitem[Young et al.(1983)]{young83}
    Young, A., Klimke, A., Africano, J. L., Quigley, R.,
    Radick, R. R., Van Buren, D.
    1983, \apj, 267, 655

\bibitem[Young et al.(1988)]{young88}
    Young, A., Skumanich, A., Paylor, V.
    1988, \apj, 334, 397



\end{thebibliography}
\end{document}